\documentstyle[pre,aps]{revtex}
\title{A Fluid Particle Model}
\author{Pep Espa\~{n}ol\footnote{e-mail: pep@fisfun.uned.es}}
\address{Departamento de F\'{\i}sica Fundamental, UNED,
Apartado 60141, 28080 Madrid, Spain}
\date{\today}

\begin{document}
\maketitle

\begin{abstract}
We present a mechanistic model for a Newtonian fluid called fluid
particle dynamics. By analyzing the concept of ``fluid particle'' from
the point of view of a Voronoi tessellation of a molecular fluid, we
propose an heuristic derivation of a dissipative particle dynamics
algorithm that incorporates shear forces between dissipative
particles. The inclusion of these non-central shear forces requires
the consideration of angular velocities of the dissipative particles
in order to comply with the conservation of angular momentum. It is
shown that the equilibrium statistical mechanics requirement that the
linear and angular velocity fields are Gaussian is sufficient to
construct the random thermal forces between dissipative particles. The
proposed algorithm is very similar in structure to the (isothermal)
smoothed particle dynamics algorithm. In this way, this work
represents a generalization of smoothed particle dynamics that
incorporates consistently thermal fluctuations and exact angular
momentum conservation. It contains also the dissipative particle
dynamics algorithm as a special case. Finally, the kinetic theory of
the dissipative particles is derived and explicit expressions of the
transport coefficients of the fluid in terms of model parameters are
obtained. This allows to discuss resolution issues for the model.
\end{abstract}

\section{Introduction}
Complex fluid systems such as colloidal or polymeric suspensions,
micelles, immiscible mixtures, etc. represent a challenge for
conventional methods of simulation due to the presence of disparate
time scales in their dynamics. There is presently a great effort in
developing new techniques of simulation that overcome some of the
difficulties of microscopic (molecular dynamics) and macroscopic
(numerical solution of continuum equations) conventional techniques.

On one hand, molecular dynamics captures all the detailed dynamics
from times scales of the order of atomic collision times to
macroscopic hydrodynamic times. However, in order to
explore these large macroscopic times the number of particles required
is enormous. Although large scale molecular dynamics simulations with
millions of particles are currently performed one realizes that not
all the information generated is actually required or even relevant at
the time scales at which rheological processes in complex fluids take
place.

On the other hand, from a continuum point of view the conventional
solution of partial differential equations like the Navier-Stokes
equation encounters difficulties due to the cumbersome treatment of
moving boundary conditions to be imposed in a system as, for example,
a colloidal suspension. These problems can be alleviated by the use of
Lagrangian descriptions in which the discretizing grid moves according
to the flow. A particularly exciting development has been the
technique of smoothed particle dynamics (SPD) which is essentially a
discretization by weight functions that transforms the partial
differential equations of continuum mechanics into ordinary
differential equations \cite{luc77,mon92}. These equations can
be further interpreted as the equations of motion for a set of
particles interacting with prescribed laws of force. The technique
thus allow to solve PDE's with molecular dynamics codes. Another
advantage of a Lagrangian description relies on the fact that no
expensive recalculations of the mesh are required as the dynamics
takes care of it. Smoothed particle dynamics has been used for
simulating astrophysical flows with non-viscous terms (this was the
original aim of the technique at the early 70's \cite{mon92}), and
very recently in the study of viscous \cite{tak94,pos95} and
thermal flows \cite{kum95,hoo96} in simple geometries.  It has
not been applied to the study of complex fluids and this may be due,
in part, to the fact that there is no easy implementation of
Lagrangian {\em fluctuating} hydrodynamics \cite{lan59} with SPD. Such
implementation would be highly desirable in order to study the
Brownian realm in which many of the processes in complex fluids take
place. It must be noted that the particulate nature of the algorithm
in SPD produces fluctuations which, from a computational point of
view, are regarded as {\em numerical noise}. It is not clear that in
the presence of viscous terms this noise satisfies the appropriate
fluctuation-dissipation theorem. A second problem with SPD is that for
viscous problems the non-central nature of the viscous shear force
breaks the conservation of angular momentum, even though the initial
continuum equations are perfectly isotropic and conserve angular
momentum.

In between microscopic and macroscopic descriptions, mesoscopic levels
of description are gaining attention in order to address flow problems
in complex fluids and/or geometries.  Lattice gas automata
\cite{fri86,wol86}, lattice Boltzmann automata \cite{lad94} or the
direct simulation Monte Carlo method for dilute gases \cite{bir76}
have been useful tools in studying hydrodynamic problems in complex
geometries. For the case of colloidal suspensions, lattice Boltzmann
techniques represent a serious competitor to Brownian dynamics
\cite{erm78} or Stokesian dynamics \cite{bra88} in that the
computational cost scales linearly with the number of colloidal
particles whereas, as a consequence of the long ranged hydrodynamic
interactions, it increases with the cube of the number of particles in
the latter techniques \cite{lad94}. A drawback of the lattice
approaches is that the dynamics is constrained by the lattice.  This
makes the consideration of boundary conditions on shaped bodies
cumbersome.

In the same spirit of looking at mesoscopic descriptions, a very
appealing idea was introduced by Hoogerbrugge and Koelman
\cite{hoo92,koe93} in which a coarse grained description of the
solvent fluid in terms of dissipative particles was devised. The
technique was coined dissipative particle dynamics (DPD) and it is an
off-lattice technique that does not suffer from the above mentioned
drawbacks of lattice gas and lattice Boltzmann simulations. DPD
consists essentially on a molecular dynamics simulation in which the
force between particles has, in addition to a conservative part, a
dissipative part represented as a Brownian dashpot. This Brownian
dashpot damps out the {\em relative approaching velocity} between
particles and introduces a noise term that keeps the system in thermal
agitation.  The dissipative particles are understood as ``droplets''
or cluster of molecules that interact with each other conserving the
total momentum of the system \cite{hoo92,esp97b}. This global
conservation law has its local counterpart in the form of a balance
equation for the momentum density, and the dissipative particles
behave hydrodynamically in the low wave number and frequency regime.

DPD has received a substantial theoretical support. It has been shown
that the original DPD algorithm of Ref. \cite{hoo92} has associated,
under a slight modification, a Fokker-Planck equation with Gibbs
equilibrium states \cite{esp95}. The extension to multicomponent
systems has also been considered \cite{cov96}. A first principles
derivation of DPD for a harmonic chain has been presented in
\cite{esp96}. The macroscopic hydrodynamic equations have been
obtained with projection operator techniques \cite{esp95a}.  A very
important further step has been the formulation of the kinetic theory
for DPD by Marsh, Backx, and Ernst \cite{mar97}, which allows to relate the
transport coefficients in the hydrodynamic equations with the DPD
model parameters. Finally, the effect of finite time steps on the
equilibrium state of the system has been considered in
\cite{mar97y}. DPD has been since applied to the study of colloidal
suspensions \cite{koe93,boe97,boe97b}, porous flow \cite{hoo92},
polymer suspensions \cite{sch95}, and multicomponent flows
\cite{cov97}.

In this work we provide a more precise meaning to the concept of
``droplet'' or ``fluid particle'' from a Voronoi tessellation of
physical space.  This conceptual framework allows to model the
different processes that intervene in the interaction between fluid
particles or mesoscopic clusters of atoms of the fluid. The outcome is
a generalization of the algorithm of DPD that includes shearing forces
between the fluid particles. These forces are non-central and do not
conserve total angular momentum. This enforces the inclusion in the
model of a spin variable with a well-defined physical meaning for the
fluid particles. In this way angular momentum conservation is
restored. We also investigate the structure of the random forces that
must be included in order to recover a Gaussian distribution of linear
and angular velocities for the fluid particles (note that the
equilibrium fluctuations of the hydrodynamic velocity {\em field} are
Gaussian). The structure of the random forces is postulated after
analogy with the structure of the random stress tensor in terms of the
Wiener process in the fluctuating hydrodynamics theory \cite{esp97}.

We show that the form of the equations of this fluid particle model at
zero temperature (when fluctuations are absent) and with no angular
variables is identical to the form of the equations obtained in a
simple version of smoothed particle dynamics as applied to fluid
systems. In this sense, this work can be regarded as a generalization
of SPD that includes fluctuations consistent with the principles of
statistical mechanics and that conserves exactly the total angular
momentum of the system. In other words, the obtained fluid particle
algorithm may be viewed as a Lagrangian discretization of the
equations of (isothermal) fluctuating hydrodynamics.

The paper is structured as follows. Section II considers the
definition of a fluid particle in terms of the Voronoi tessellation
and this serves to motivate the type of forces and torques between
fluid particles introduced in Section III. Section IV presents
the equivalent Fokker-Planck equation to the equations of motion.
This allows to establish requirements on the model parameters in
order to have a proper equilibrium distribution. A summary of
the model is presented in Section V. Section VI contains
the kinetic theory of the model in the simple case when conservative
forces are absent. The transport coefficients are computed and
this permits to discuss its dependence on the number density of
fluid particles in Section VII. A final discussion and conclusions
is presented in the last section.

\section{Fluid particles through Voronoi tessellation}
In an attempt to better understand the physical meaning of DPD, we
have devised a coarse graining procedure for a molecular dynamics
simulation of point particles (atoms) interacting through continuous
potentials \cite{esp97b}. The coarsening is performed through the
Voronoi tessellation that allows to divide physical space into a set
of non-overlapping cells that cover all the space in a well-defined
manner. Given a discrete set of points (that can be distributed at
random) the Voronoi tessellation assigns to each point (called
``cell center''), that region of space that surrounds it and that is
nearer to this point than to any other point of the set. With
this tessellation the atoms of the molecular dynamics simulation are
distributed into clusters around the centers of the Voronoi cells. The
practical way to perform the Voronoi tessellation in the simulation is
by computing the distance of a given atom to all the center cells and
assign that atom to the nearest center. Subsequently, the Voronoi
cells are set in motion according to the velocity and acceleration of
the center of mass of the particles that are within the cell. In this
way, the cells capture the motion of the fluid at mesoscopic scales.

The Voronoi cells are a well defined realization of what is loosely
regarded in fluid mechanics textbooks as ``fluid particles''. We would
like to know how these fluid particles move, that is, which explicit
law of force between fluid particles would reproduce the actual motion
of the Voronoi cells observed in the simulations. It is apparent that
the number of cells is much smaller than the number of atoms in the
molecular dynamics simulation, and therefore if one knows how the
clusters move, one can try to simulate the clusters and capture the
mesoscopic behavior of the underlying liquid with much less
computational effort.

For sufficiently large clusters, that is, when the typical distance
$\lambda_{mes}$ between cell centers is much larger than the typical
distance $\lambda_{mic}$ between atoms we expect that the clusters
move {\em hydrodynamically}. More precisely, we consider the
conserved densities

\begin{eqnarray}
\rho_{\bf r}&=&\frac{M_{\bf r}}{V_{\bf r}}
\nonumber\\
{\bf g}_{\bf r}&=&\frac{{\bf P}_{\bf r}}{V_{\bf r}}
\nonumber\\
e_{\bf r}&=&\frac{E_{\bf r}}{V_{\bf r}}
\label{densit}
\end{eqnarray}
where $M_{\bf r},{\bf P}_{\bf r}, E_{\bf r}$ are the instantaneous
total mass, momentum and energy, respectively, of the system of
particles that happen to be within the Voronoi cell centered at ${\bf
r}$ and $V_{\bf r}$ is the volume of the cell. One also introduces

\begin{equation}
{\bf v}_{\bf r}\equiv\frac{{\bf g}_{\bf r}}{\rho_{\bf r}}
\label{e6}
\end{equation}
which is the instantaneous velocity of the center of mass of the
system of particles within the Voronoi cell at ${\bf r}$.

If, 1) the cells are large enough for the system of particles that are
within it to be considered as a thermodynamic system, and 2) the
variations of the conserved quantities from neighbor cells are small,
then the variables (\ref{densit}) obey the equations of
fluctuating hydrodynamics \cite{lan59} (see \cite{esp97} for
the non-linear case). The conserved quantities
(\ref{densit}) are subject to fluctuations because the atoms can enter
and go out from the Voronoi cells due to their thermal agitation. The
size of fluctuations, that is, the noise amplitude appearing in the
equations of fluctuating hydrodynamics is proportional to the square
root of the inverse of the volume of the cell, in accordance with the
$1/\sqrt{N}$ dependence of fluctuations in equilibrium ensemble
theory \cite{esp97}. Therefore, depending on the ``resolution'' (the
number of Voronoi cells per unit volume) used to describe the system,
the amplitude of the noise term in the hydrodynamic equations that
govern the instantaneous values of the conserved variables will be
different.

Now, one is faced with two possible routes in order to simulate the
dynamics of clusters. The first route is to consider the conserved
discrete variables (\ref{densit}) as the state variables and update
them according to some discretized version of the equations of
hydrodynamics.  This poses some subtle problems regarding the
formulation of {\em fluctuating} hydrodynamics in a moving mesh, in
particular with the treatment of the $1/\sqrt{V_{\bf r}}$
singularity. The second route, which is the one we follow in this
paper, is to {\em postulate} the laws of force between cells. Despite
the strong assumptions made to model these forces, the final
expressions satisfy symmetry requirements that ensure that the
behavior of the clusters will be, on average, that of real clusters.

A final word on the angular momentum is in order.
We have not included in the above set of conserved variables
(\ref{densit}) the angular momentum density defined by

\begin{equation}
{\bf J}_{{\bf r}}={\bf r}\times {\bf g}_{{\bf r}}+\frac{{\bf S}_{{\bf r}}}
{v_{{\bf %
r}}}  
\label{ang1}
\end{equation}
where we have decomposed the angular momentum of the system of particles in
cell ${\bf r}$ as the sum of the angular momentum of the center of mass with
respect to the origin plus the intrinsic angular momentum ${\bf S}_{{\bf r}}$
of the particles of the cell with respect to the center of mass of the cell. We
could, in principle, compute the inertia tensor ${\bf I}_{{\bf r}}$ for the
system of particles in cell ${\bf r}$ and define the angular velocity ${\bf %
\omega }_{{\bf r}}$ through 
\begin{equation}
{\bf \omega }_{{\bf r}}={\bf I}_{{\bf r}}^{-1}{\bf S}_{{\bf r}}
\label{omega}
\end{equation}

The reason why the angular momentum density is usually not considered
in the derivation of Newtonian hydrodynamics is because the second
contribution in (\ref{ang1}) vanishes in the ``continuum limit''. This
can be seen by noting that ${\bf S_r}$ must scale as a typical size of
the cell. In the continuum limit this typical distance goes to zero
and there is no intrinsic angular momentum contribution. The situation
here is different from the case of molecular fluids with spin
\cite{gro62}, where the rotation of the molecules themselves
originates an angular momentum that does not scale with the size of
the cells and produces a finite value in the continuum limit.

\section{Modeling the forces and torques between fluid particles}
In this section we formulate the fluid particle model under a set of
simplifying hypothesis.  In the real clusters, the mass is a
fluctuating quantity as particles can enter and go out from the
Voronoi cell. Also, the shape of the cells changes as the cells
move. However, the first assumption is that all clusters are
identical, having a fixed mass $m$ and fixed isotropic inertia tensor
of moment of inertia $I$. We assume that the state of the cluster
system is completely characterized by the positions ${\bf r}_i$, the
velocities of the center of mass ${\bf v}_i$ and the angular
velocities ${\bf \omega}_i$. Note that we do not include any internal
energy variable and therefore, the resulting algorithm will not
capture appropriately the thermal effects that occur in real
fluids. This may be a minor problem when one is interested only in
rheological properties. A generalization of the model including energy
conservation has been recently proposed independently in Refs. \cite{bon97}
\cite{esp-ener97}.

The next step is to specify the forces and torques that are
responsible for changing the values of the linear and angular
velocities of the clusters. We model the forces between two clusters
by considering several heuristic arguments about how one expects that
the actual Voronoi clusters interact with each other. In this respect
we make first a strong pair-wise additivity assumption. In the real
system one expects that the force between two clusters (that is
between all the atoms of the first cluster that are interacting with
the atoms of the second cluster) will depend in general not only on
the state variables of these two clusters but also on the
configuration of other neighboring clusters. For the sake of simplicity,
though, we neglect this collective effect and assume that the force
between two clusters depends only on the position and velocities of
these two clusters.

The equations of motion are therefore

\begin{eqnarray}
\dot{{\bf r}}_i&=&{\bf v}_i
\nonumber\\
\dot{{\bf v}}_i&=&\frac{1}{m}\sum_{j\neq i}{\bf F}_{ij}
\nonumber\\
\dot{{\bf \omega}}_i&=&\frac{1}{I}\sum_{j\neq i}{\bf N}_{ij}
\label{eqmotion}
\end{eqnarray}
where ${\bf F}_{ij},{\bf N}_{ij}$ are the force and torque that
cluster $j$ exerts on cluster $i$.  We require that the forces
satisfy Newton's third law, ${\bf F}_{ij}=-{\bf F}_{ji}$, in such
a way that the total linear momentum ${\bf P}=\sum_i m {\bf v}_i$ is
a dynamical invariant, $\dot{\bf P}=0$. In addition, we assume
that the torques in (\ref{eqmotion}) are given by
\begin{equation}
{\bf N}_{ij}=-\frac{1}{2}{\bf r}_{ij}\times{\bf F}_{ij}
\label{m1}
\end{equation}
and one checks immediately that the total angular momentum 

\begin{equation}
{\bf J}=\sum_i{\bf r}_i\times{\bf p}_i + I\omega_i
\label{tm}
\end{equation}
is conserved exactly, $\dot{{\bf J}}=0$.

We will model the force between clusters according to

\begin{equation}
{\bf F}_{ij}={\bf F}^C_{ij}+{\bf F}^T_{ij}+{\bf F}^R_{ij}
+{\tilde{\bf F}}_{ij}
\label{f1}
\end{equation}
The first three contributions are deterministic forces whereas the
last one is random. Let us discuss them separately.

\subsection{Deterministic forces}
The first contribution ${\bf F}^C_{ij}$ to the force is assumed to
arise from a conservative potential $V(r)$ that depends on the
separation distance between clusters. In Ref. \cite{esp97b} we have
argued that a plausible definition of this potential is through the
logarithm of the radial distribution function of cluster centers. The resulting soft
potential has a bell-shaped form and has the virtue that when used as
the potential between clusters in a MD simulation, it reproduces
consistently the radial distribution function of the real clusters (as it has
been checked through an actual MD simulation). It
therefore captures the static or equilibrium properties of the system
of clusters. The physical interpretation of this force is that it
provides the excluded volume effect of each cluster. The center of the
cluster (and its center of mass) is usually located ``in the middle''
of the cell and therefore it is not very probable that two cluster
centers are closer to each other than the typical size of the cell.

It is clear, though, that this conservative contribution
cannot be the only contribution to the force because it does not
capture friction effects between clusters. These friction effects will
depend on the velocities between clusters and will give rise to
dissipative processes. The second contribution in (\ref{f1}) is a
friction force that depends on the relative translational velocities
of the clusters $i,j$ with positions ${\bf r}_i,{\bf r}_j$ and
velocities ${\bf v}_i,{\bf v}_j$ in the following way

\begin{equation}
{\bf F}_{ij}^T= -\gamma m {\bf M}^T({\bf r}_{ij})\!\cdot\!{\bf v}_{ij}
\label{dis1}
\end{equation}
where ${\bf v}_{ij}={\bf v}_i-{\bf v}_j$ is the relative velocity and
the dimensionless matrix ${\bf M}^T({\bf r}_{ij})$ is the most general
matrix that can be constructed out of the vector ${\bf r}_{ij}={\bf
r}_i-{\bf r}_j$, this is

\begin{equation}
{\bf M}^T({\bf r}_{ij})
\equiv A(r_{ij}){\bf 1}+B(r_{ij}){\bf e}_{ij}{\bf e}_{ij}
\label{mt}
\end{equation}
where ${\bf 1}$ is the unit matrix, ${\bf e}_{ij}={\bf r}_{ij}/r_{ij}$
is the unit vector joining the particles, $r_{ij}=|{\bf r}_{ij}|$ and
the dimensionless functions $A(r)$ and $B(r)$ provide the range of the
force. The friction coefficient $\gamma$ has been introduced as an
overall factor for convenience and has dimensions of inverse of
time. The first contribution to the dissipative force (\ref{dis1}) is
in the direction of the relative velocities and tends to damp out the
difference between the velocities. {\em It is a shearing force which
is non-central}. The second contribution is directed along the joining
line of the particles and damps out the relative approaching motion of
the particles. The dissipative force in the original algorithm of
Hoogerbrugge and Koelman is obtained with $A(r)=0$ \cite{hoo92}. Note
that the form of the force ${\bf F}_{ij}^T$ is the more general
expression for a vector that depends on ${\bf r}_{ij}$ and is linear
in the relative velocities.

We now discuss the effects of rotation in the dissipative force. Let
us assume for a moment that the clusters $i$ and $j$ were spheres of
radius $r_{ij}/2$ in contact and spinning with angular velocities
${\bf \omega}_i,{\bf \omega}_j$ with no translational velocities. We
would have a relative velocity at the ``surface'' of the spheres equal
to $\frac{1}{2}{\bf r}_{ij}\times({\bf \omega}_i+{\bf \omega}_j)$ and
it is plausible to associate a friction force between the spheres
proportional (in matrix sense) to this relative velocity. Therefore,
the rotational contribution to the dissipative force is of the form

\begin{equation}
{\bf F}^R_{ij}= -\gamma m {\bf M}^R({\bf r}_{ij}) \!\cdot\!
\left(\frac{{\bf r}_{ij}}{2}\times[{\bf \omega}_i+{\bf \omega}_j] \right) 
\label{rot}
\end{equation}
Again, the dimensionless matrix ${\bf M}^R$ depends only on the vector
${\bf r}_{ij}$ and therefore it must have the form

\begin{equation}
{\bf M}^R({\bf r}_{ij})
=C(r_{ij}){\bf 1}
+D(r_{ij}){\bf e}_{ij}{\bf e}_{ij}
\label{simp}
\end{equation}
where $C(r),D(r)$ are scalar functions. The first part of the matrix
gives rise to a friction force proportional to the relative velocity
at the ``surface'' of the spheres. The effect of this force is
twofold.  On one hand, the spinning of a particle causes translational
motion onto the neigbouring particles. On the other, it also causes
rotational motion in such a way that two neigbouring particles prefer
to be with opposite angular velocities (in a sort of ``engaging''
effect). The presence of a third particle frustrates the spinning of
both particles and, therefore, the global effect of this force is to
damp out to zero the angular velocities of the particles. The second
contribution to the force (\ref{simp}) is actually zero because the
cross product is perpendicular to ${\bf e}_{ij}$. We retain this term
just to maintain the analogy between both matrices ${\bf M}^R$ in
(\ref{dis1}) and ${\bf M}^T$ in (\ref{rot}). Finally, we use the same
value for $\gamma$ in (\ref{dis1}) and in (\ref{simp}) because any
difference can be taken into account through the functions
$A(r),B(r),C(r), D(r)$.

If we use, instead of the polar vector representation for the angular
velocity, the antisymmetric tensor representation ${\bf
\varpi}_{xy}=-{\bf \varpi}_{yx}={\bf \omega}_z$ (cyclic), we can write
the force in the form

\begin{equation}
{\bf F}_{ij}^{R}=-\gamma m
{\bf M}^R({\bf r}_{ij})\!\cdot\!
({\bf \varpi}_{i}+
{\bf \varpi }_{j})\!\cdot \!\frac{{\bf r}_{ij}}{2}
\label{rot2}
\end{equation}
which shows explicitly the vectorial nature of the force (this is, 
${\bf F}_{ij}^{R}$ transforms under rotations as a vector). 

\subsection{Random forces}
The first three contributions ${\bf F}^C_{ij},{\bf F}^T_{ij},{\bf
F}^R_{ij}$ to the force between clusters in (\ref{f1}) are
deterministic while the last one is stochastic. The reason why we
introduce a random force is because it is well-known that whenever a
coarse-graining procedure is performed, dissipation and noise arise
and both are related through a fluctuation-dissipation theorem. After
discussing the form of the dissipative forces we will now consider the
form that the random force should have.

Inspired by the tensorial structure of the random forces that appear
in the fluctuating hydrodynamics theory \cite{esp97}, we expect that
the dissipation due to shear has associated a traceless symmetric
random matrix and that the dissipation due to compressions has
associated a diagonal trace matrix. By symmetry reasons, we expect
that the noise associated to rotational dissipation will involve an
antisymmetric matrix. Therefore, we {\em postulate} the following
velocity independent random force

\begin{equation}
{\tilde{\bf F}}_{ij} dt =\sigma m
\left({\tilde A}(r_{ij}){\overline{d{\bf W}}}^S_{ij}
+{\tilde B}(r_{ij})\frac{1}{D}{\rm tr}[d{\bf W}_{ij}]{\bf 1}
+{\tilde C}(r_{ij}){d{\bf W}}^A_{ij}\right)
\!\cdot\!{\bf e}_{ij}
\label{ran1}
\end{equation}
where ${\tilde A}(r),{\tilde B}(r),{\tilde C}(r)$ are scalar
functions, $\sigma$ is a parameter governing the overall noise
amplitude, and we introduce the following symmetric, antisymmetric and
traceless symmetric random matrices
\begin{eqnarray}
d{\bf W}^{S\mu\nu}_{ij}
&=&\frac{1}{2}\left[d{\bf W}^{\mu\nu}_{ij}+d{\bf W}^{\nu\mu}_{ij}\right]
\nonumber \\
d{\bf W}^{A\mu\nu}_{ij}
&=&\frac{1}{2}\left[d{\bf W}^{\mu\nu}_{ij}-d{\bf W}^{\nu\mu}_{ij}\right]
\nonumber \\
\overline{d{\bf W}}^S_{ij}&=&d{\bf W}^S_{ij}-\frac{1}{D}{\rm
tr}[d{\bf W}^S_{ij}]{\bf 1}
\label{decomp}
\end{eqnarray}
The overline in a matrix denotes its traceless part.
Here, $D$ is the physical dimension of space and $d{\bf
W}^{\mu\nu}_{ij}$ is a matrix of independent Wiener increments which is
assumed to be symmetric under particle interchange
\begin{equation}
d{\bf W}^{\mu\nu}_{ij}=d{\bf W}^{\mu\nu}_{ji}
\end{equation}
This symmetry will ensure momentum conservation because $\tilde{\bf
F}_{ij}=-\tilde{\bf F}_{ji}$.  The matrix $d{\bf W}^{\mu\nu}_{ij}$ is
an infinitesimal of order $1/2$, and this is summarized in the Ito
mnemotechnical rule
\begin{equation}
d{\bf W}^{\mu\mu'}_{ii'}d{\bf W}^{\nu\nu'}_{jj'}=
\left[\delta_{ij}\delta_{i'j'}+\delta_{ij'}\delta_{ji'}\right]
\delta_{\mu\nu}\delta_{\mu'\nu'}dt
\label{ran3}
\end{equation}
From this stochastic property, one derives straightforwardly the
following rules from the different parts in (\ref{ran1})

\begin{eqnarray}
{\rm tr}[d{\bf W}_{ii'}]{\rm tr}[d{\bf W}_{jj'}]
&=&\left[\delta_{ij}\delta_{i'j'}+\delta_{ij'}\delta_{ji'}\right]
Ddt
\nonumber\\
\overline{d{\bf W}}^{S\mu\mu'}_{ii'}
\overline{d{\bf W}}^{S\nu\nu'}_{jj'}
&=&\left[\delta_{ij}\delta_{i'j'}+\delta_{ij'}\delta_{ji'}\right]
\left[
\frac{1}{2}\left(\delta_{\mu\nu}\delta_{\mu'\nu'}+
\delta_{\mu\nu'}\delta_{\mu'\nu}\right)-\frac{1}{D}
\delta_{\mu\mu'}\delta_{\nu\nu'}
\right]dt
\nonumber\\
d{\bf W}^{A\mu\mu'}_{ii'}
d{\bf W}^{A\nu\nu'}_{jj'}&=&
\left[\delta_{ij}\delta_{i'j'}+\delta_{ij'}\delta_{ji'}\right]
\frac{1}{2}\left(\delta_{\mu\nu}\delta_{\mu'\nu'}-
\delta_{\mu\nu'}\delta_{\mu'\nu}\right)dt
\nonumber\\
{\rm tr}[d{\bf W}_{ii'}]d\overline{{\bf W}}^S_{jj'}&=&0
\nonumber\\
{\rm tr}[d{\bf W}_{ii'}]d{\bf W}^A_{jj'}&=&0
\nonumber\\
\overline{d{\bf W}}^{S\mu\mu'}_{ii'}
d{\bf W}^{A\nu\nu'}_{jj'}&=&0
\label{ran4}
\end{eqnarray}
These expressions show that the traceless symmetric, the trace and the
antisymmetric parts are independent stochastic processes. The
apparently complex structure of the random force (\ref{ran1}) is
required in order to be consistent with the tensor structure of the
dissipative friction forces (\ref{dis1}) and (\ref{rot}). This will
become apparent when considering the associated Fokker-Planck equation
in the next section and requiring that it has a proper equilibrium
ensemble.

Despite of the heuristic arguments and strong assumptions made in
order to model the force ${\bf F}_{ij}$ between clusters, we note that
this force is the most general force that can be constructed out of
the vectors ${\bf r}_i,{\bf r}_j,{\bf v}_i,{\bf v}_j,{\bf
\omega}_i,{\bf \omega}_j$ and that satisfies the following properties:

\begin{enumerate}
\item It is invariant under translational and Galilean transformations
and transforms as a vector under rotations.
\item It is linear in the linear and angular velocities. This
linearity is required in order to be consistent with the Gaussian
distribution of velocities at equilibrium, as we will show later.
\item It satisfies Newton's third law ${\bf F}_{ij}=-{\bf F}_{ji}$
and, therefore, the total linear momentum will be a conserved quantity
of the system.
\end{enumerate}

\section{Fokker-Planck equation and equilibrium state}
The equations of motion (\ref{eqmotion}) are Langevin equations which
in the form of proper stochastic differential equations (SDE) become

\begin{eqnarray}
d{\bf r}_i &=& {\bf v}_i dt
\nonumber\\
d{\bf v}_i &=&\frac{1}{m} 
\sum_{i'}\left[{\bf F}^C_{ii'}+{\bf F}^T_{ii'}+{\bf F}^R_{ii'}\right]dt
+\sum_{i'}d{\tilde{\bf v}}_{ii'}
\nonumber\\
d{\bf \omega}_i &=&
-\frac{1}{I} 
\sum_{i'}\frac{{\bf r}_{ii'}}{2}
\times\left[{\bf F}^T_{ii'}+{\bf F}^R_{ii'}\right]dt
-\frac{m}{I} \sum_{i'}\frac{{\bf r}_{ii'}}{2} \times d{\tilde{\bf v}}_{ii'}
\label{sdea}
\end{eqnarray}
where we have introduced 

\begin{equation}
 d{\tilde{\bf v}}_{ii'} \equiv\frac{1}{m} \tilde{{\bf F}}_{ii'} dt 
=\sigma\left({\tilde A}(r_{ii'}){\overline{d{\bf W}}}^S_{ii'}
+{\tilde B}(r_{ii'})\frac{1}{D}{\rm tr}[d{\bf W}_{ii'}]{\bf 1}
+{\tilde C}(r_{ii'})d{\bf W}^A_{ii'}\right)\!\cdot\!{\bf e}_{ii'}
\label{dv}
\end{equation}
In principle, one should specify which stochastic interpretation
(It\^o or Stratonovich) must be used in these equations
\cite{gar83}. Nevertheless, both interpretations produce the same
answers because the random forces are velocity independent.

Associated to the SDE (\ref{sdea}) there exists a mathematically
equivalent Fokker-Planck equation (FPE).  The FPE governs the
distribution function $\rho( r, v,\omega; t)$ that gives the
probability density that the $N$ clusters of the system have specified
values for the positions, velocities and angular velocities. 
We show in the  Appendix that the FPE is given by

\begin{equation}
\partial_t \rho(r,v,\omega;t)=\left[L^C+L^T+L^R\right]\rho(r,v,\omega;t)
\label{fp1}
\end{equation}
The operator $L^C$ is the usual Liouville operator of a Hamiltonian
system interacting with conservative forces ${\bf F}^C$,
this is,
\begin{equation}
L^C=-\left[\sum_i{\bf v}_i\frac{\partial }{\partial {\bf r}_i}
+\sum_{i,j\neq i}\frac{1}{m}{\bf F}^C_{ij}
\frac{\partial}{\partial{\bf v}_i}\right]
\label{liouvilleop}
\end{equation}
The operators
$L^T,L^R$ are given by
\begin{eqnarray}
L^T&=&
\sum_{i,j\neq i}\frac{\partial}{\partial {\bf v}_i}
\!\cdot\!\left[{\bf L}^T_{ij}+{\bf L}^R_{ij}\right]
\nonumber \\
L^R&=&-\frac{m}{I}
\sum_{i,j\neq i}\frac{\partial}{\partial {\bf \omega}_i}
\!\cdot\!\left(\frac{{\bf r}_{ij}}{2}\times
\left[{\bf L}^T_{ij}+{\bf L}^R_{ij}\right]\right)
\label{fp3}
\end{eqnarray}
with
\begin{eqnarray}
{\bf L}^T_{ij}
&\equiv&-\frac{1}{m}{\bf F}^T_{ij}
+\frac{\sigma^2}{2}{\bf T}_{ij}\!\cdot\!
\left[\frac{\partial}{\partial{\bf v}_i}-\frac{\partial}{\partial{\bf v}_j}\right]
\nonumber\\
{\bf L}^R_{ij}&\equiv&
-\frac{1}{m}{\bf F}^R_{ij}
+\frac{m}{I}\frac{\sigma^2}{2}{\bf T}_{ij}
\!\cdot\!
\left( \frac{{\bf r}_{ij}}{2}\times \left[\frac{\partial}{\partial{\bf \omega}_i}
+\frac{\partial}{\partial{\bf \omega}_j}\right]\right)
\label{vecoper}
\end{eqnarray}
Here, the matrix ${\bf T}_{ij}$ is given by
\begin{equation}
{\bf T}_{ij}=
\frac{1}{2}\left[{\tilde A}^2(r_{ij})
+{\tilde C}^2(r_{ij})\right]{\bf 1}
+\left[\left(\frac{1}{2}-\frac{1}{D}\right)
{\tilde A}^2(r_{ij})+\frac{1}{D}{\tilde B}^2(r_{ij})
-\frac{1}{2}{\tilde C}^2(r_{ij})\right]
{\bf e}_{ij}{\bf e}_{ij}
\label{t}
\end{equation}

The steady state solution of equation (\ref{fp1}), $\partial_t
\rho=0$, gives the equilibrium distribution $\rho^{eq}$. We now
consider the conditions under which the steady state solution is the
Gibbs canonical ensemble:

\begin{eqnarray}
\rho^{eq}(r,v,\omega)&=&\frac{1}{Z}\exp\{-H(r,v,\omega)/k_BT\}
\nonumber \\
&=&\frac{1}{Z}\exp\{-\left(\sum_i\frac{m}{2}v_i^2+
\frac{I}{2}\omega_i^2+V(r)\right)/k_BT\}
\label{eq}
\end{eqnarray}
where $H$ is the Hamiltonian of the system, $V$ is the potential
function that gives rise to the conservative forces ${\bf F}^C$, $k_B$
is Boltzmann's constant, $T$ is the equilibrium temperature and $Z$ is
the normalizing partition function. We note that the velocity and
angular velocity {\em fields} are Gaussian variables at equilibrium
and, therefore, one expects that the distribution function of the
discrete values of theses fields is also Gaussian.

The canonical ensemble is the
equilibrium solution for the conservative system, {\it i.e.}
$L^C\rho^{eq}=0$. If in addition the following equations are
satisfied
\begin{eqnarray}
{\bf L}^T_{ij}\rho^{eq}&=&0
\nonumber\\
{\bf L}^R_{ij}\rho^{eq}&=&0
\label{restric}
\end{eqnarray}
then we will have $L\rho^{eq}=0$ and the Gibbs equilibrium
ensemble will be the unique stationary  solution of the dynamics.
Eqns. (\ref{restric}) will be satisfied if 

\begin{equation}
\gamma =\frac{\sigma^2m}{2k_BT}
\label{db1}
\end{equation}
which is a detailed balance condition, and also
\begin{equation}
{\bf M}^R({\bf r}_{ij})={\bf M}^T({\bf r}_{ij})={\bf T}_{ij}
\label{mat}
\end{equation}
This is the fluctuation-dissipation theorem for the fluid particle model.
We observe, therefore, that the initial hypothesis for the tensorial
structure of the dissipative (\ref{dis1}), (\ref{rot2}) and random
(\ref{ran1}) forces was correct and consistent with equilibrium
statistical mechanics.

A final word about an H-theorem is in order. It has been shown in
Ref. \cite{mar97} that the original DPD algorithm has an H-theorem
that ensures that the equilibrium ensemble is the final solution for
whatever initial condition selected. In the model presented in this
paper there is also a functional of $\rho(z)$ that is a Lyapunov
functional. It is not necessary to prove again that an H-theorem
exists for the fluid particle model, because a {\em
general} H-theorem exists for {\em any} Fokker-Planck equation
\cite{ris84}. The only condition is that the diffusion matrix
accompanying the second derivative terms of the FPE is positive (semi)
definite.  However, in the model presented in this paper the diffusion
matrix is positive semidefinite by construction, because the FPE has
been obtained from a SDE. The diffusion matrix is obtained from the
product of two identical matrices. Therefore, its eigenvalues are the
square of the eigenvalues of these matrices and are necessarily
positive (or zero).

\section{Summary of the fluid particle model}
In this section and for the sake of clarity we collect the results
presented so far. The fluid particle model is defined by $N$ identical
particles of mass $m$ and moment of inertia $I$. The state of the
system is characterized by the positions ${\bf r}_i$, velocities ${\bf
v}_i$, and angular velocities ${\bf \omega}_i$ of each particle. The
forces and torques on the particles are given by
\begin{eqnarray}
{\bf F}_i &=& \sum_j {\bf F}_{ij}
\nonumber \\
{\bf N}_i &=& -\sum_j \frac{{\bf r}_{ij}}{2}\times {\bf F}_{ij}
\label{sum1}
\end{eqnarray}
where the force that particle $j$ exerts on particle $i$ is
given by

\begin{equation}
{\bf F}_{ij}={\bf F}^C_{ij}+{\bf F}^T_{ij}+{\bf F}^R_{ij}
+{\tilde{\bf F}}_{ij}
\label{sum2}
\end{equation}
The conservative ($C$), translational ($T$), rotational ($R$) and
random ($\sim$) contributions are given by

\begin{eqnarray}
{\bf F}^C_{ij} &=& -V'(r_{ij}){\bf e}_{ij}
\nonumber \\
{\bf F}^T_{ij} &=& -\gamma m {\bf T}_{ij}\!\cdot\!{\bf v}_{ij}
\nonumber \\
{\bf F}^R_{ij} &=& -\gamma m {\bf T}_{ij}\!\cdot\!
\left(\frac{{\bf r}_{ij}}{2}\times({\bf \omega}_{i}+{\bf \omega}_{j})\right)
\nonumber\\
{\tilde{\bf F}}_{ij} dt &=& \left(2k_BT\gamma m\right)^{1/2}
\left({\tilde A}(r_{ij}){\overline{d{\bf W}}}^S_{ij}
+{\tilde B}(r_{ij})\frac{1}{D}{\rm tr}[d{\bf W}_{ij}]{\bf 1}
+{\tilde C}(r_{ij}) d{\bf W}^A_{ij}\right)
\!\cdot\!{\bf e}_{ij}
\label{sum3}
\end{eqnarray}
The random bits are defined in (\ref{decomp}) and its stochastic
properties are given in (\ref{ran4}). Here, the matrix ${\bf T}$ is
given by

\begin{equation}
{\bf T}_{ij}= A(r_{ij}){\bf 1}+B(r_{ij}){\bf e}_{ij}{\bf e}_{ij}
\label{t2}
\end{equation}
where
\begin{eqnarray}
A(r)&=&\frac{1}{2}\left[{\tilde A}^2(r)+{\tilde C}^2(r)\right]
\nonumber\\
B(r)&=&
\frac{1}{2}\left[{\tilde A}^2(r)-{\tilde C}^2(r)\right]
+\frac{1}{D}\left[{\tilde B}^2(r)-{\tilde A}^2(r)\right]
\label{AB}
\end{eqnarray}

The model is thus specified by providing the scalar functions
$V(r),{\tilde A}(r),{\tilde B}(r),{\tilde C}(r)$. We note that the
case ${\tilde A}(r)={\tilde C}(r)=0$ corresponds to the original DPD
algorithm of Hoogerbrugge and Koelman \cite{hoo92,esp95}. In this
case, the random force is given in terms of a single random number
(the trace), the forces are central and the torques vanish, rendering
the spin variables unnecessary. Note that there is some freedom in
selecting the functions ${\tilde A}(r),{\tilde B}(r),{\tilde C}(r)$
and it might be convenient to take ${\tilde A}(r)$ or ${\tilde C}(r)$
equal to zero in order to compute only four of seven random numbers in
each step of a simulation.

The model presented in the language of SDE is more appropriate for its
direct use in simulations. For theoretical analysis it is more
convenient to use the corresponding FPE which is given by

\begin{equation}
\partial_t \rho(r,v,\omega;t)=\left[L^C+L^T+L^R\right]\rho(r,v,\omega;t)
\label{sum4}
\end{equation}
where 
\begin{eqnarray}
L^C &=& -\sum_i {\bf v}_i\!\cdot\!\frac{\partial}{\partial {\bf r}_i}
+
\frac{{\bf F}^C_i}{m}\!\cdot\!\frac{\partial}{\partial {\bf v}_i}
\nonumber\\
L^T&=&\sum_{i,j\neq i}\frac{\partial}{\partial {\bf v}_i}
\!\cdot\!\left[{\bf L}^T_{ij}+{\bf L}^R_{ij}\right]
\nonumber \\
L^R&=&-\frac{m}{I}
\sum_{i,j\neq i}\frac{\partial}{\partial {\bf \omega}_i}
\!\cdot\!\left(\frac{{\bf r}_{ij}}{2}
\times\left[{\bf L}^T_{ij}+{\bf L}^R_{ij}\right]\right)
\label{sum5}
\end{eqnarray}
Here, the vector operators are given by

\begin{eqnarray}
{\bf L}^T_{ij}
&\equiv&\gamma {\bf T}_{ij}\!\cdot\!
\left[{\bf v}_{ij}
+\frac{k_BT}{m}
\left[\frac{\partial}{\partial{\bf v}_i}-\frac{\partial}{\partial{\bf v}_j}\right]\right]
\equiv \gamma {\bf T}_{ij}\!\cdot\!{\cal V}_{ij}
\nonumber\\
{\bf L}^R_{ij}
&\equiv&\gamma {\bf T}_{ij}\!\cdot\!
\left[\left(\frac{{\bf r}_{ij}}{2}\times 
\left[{\bf \omega}_i+{\bf \omega}_j\right]\right)
+\frac{k_BT}{I}
\left( \frac{{\bf r}_{ij}}{2}\times \left[\frac{\partial}{\partial{\bf \omega}_i}
+\frac{\partial}{\partial{\bf \omega}_j}\right]\right)\right]
\equiv \gamma {\bf T}_{ij}\!\cdot\!{\cal W}_{ij}
\label{sum6}
\end{eqnarray}
were the last equality defines the two vector operators ${\cal
V}_{ij},{\cal W}_{ij}$.

\section{Kinetic theory}
One would like to predict the macroscopic behavior of the fluid particle
model and, in particular, check that this behavior conforms to the laws
of hydrodynamics (as expected from symmetry considerations) and predict
the value of the transport coefficients in terms of model parameters. 
The global conservation laws of mass, linear and angular momentum in
the fluid particle model have a local counterpart in the form of
balance equations. Our aim is to formulate these equations of
transport within a kinetic theory approach, as has been done by Marsh,
Backx, and Ernst recently for the case of the original DPD model in
Ref. \cite{mar97}. A derivation of the hydrodynamic equations with a
projection operator technique for the original DPD algorithm was
presented in Ref. \cite{esp95a}. The projector used was the Mori
projector \cite{mor65} and the resulting equations were the linearized
equations of hydrodynamics. By using a {\em time-dependent} projector
one can obtain the {\em non-linear} equations of hydrodynamics with
the transport coefficients expressed in terms of Green-Kubo formulae
\cite{gra82}. Although explicit calculations can be performed of these
Green-Kubo formulae under certain approximations \cite{rei89}, we
adopt in this paper the approach of kinetic theory, allowing for a
straightforward comparison with the results of Ref. \cite{mar97}.

\subsection{General rate of change equation}

The starting point is the formulation of the general rate of
change equation for an arbitrary function $G(z)$ where $z$ is
a shorthand for the set of all positions, velocities and angular
velocities of the $N$ particles of the fluid. By using the
Fokker-Planck equation (\ref{sum4}), we can write

\begin{eqnarray}
\partial_t\langle G\rangle &=&\int dz G(z)\partial_t \rho(z;t)
\nonumber\\
&=&\int dz G(z)\left[ L^C+L^T+L^R\right] \rho(z;t)
\nonumber\\
&=&\int dz  \rho(z;t)\left[ L^{C+}+L^{T+}+L^{R+}\right]G(z)
\label{rate1}
\end{eqnarray}
where an integration by parts is performed and the adjoint operators are
defined by

\begin{eqnarray}
L^{C+} &=& \sum_i {\bf v}_i\!\cdot\!\frac{\partial}{\partial {\bf r}_i}
+
\frac{{\bf F}^C_i}{m}\!\cdot\!\frac{\partial}{\partial {\bf v}_i}
\nonumber\\
L^{T+}&=&\gamma\sum_{i,j\neq i}
\left({\overline {\bf {\cal V}}}_{ij}
+{\overline {\bf {\cal W}}}_{ij}\right)
\!\cdot\!{\bf T}_{ij}\!\cdot\!
\frac{\partial}{\partial {\bf v}_i}
\nonumber \\
L^{R+}&=&-\frac{m}{I}\gamma
\sum_{i,j\neq i}
\left(\frac{{\bf r}_{ij}}{2}
\times{\bf T}_{ij}\!\cdot\!({\overline {\bf {\cal V}}}_{ij}
+{\overline {\bf {\cal W}}}_{ij})\right)\!\cdot\!
\frac{\partial}{\partial {\bf \omega}_i}
\label{rate2}
\end{eqnarray}
Here, the vector operators are given by

\begin{eqnarray}
 {\bf {\cal V}}^+_{ij}&\equiv&-{\bf v}_{ij}
+\frac{k_BT}{m}
\left[\frac{\partial}{\partial{\bf v}_i}
-\frac{\partial}{\partial{\bf v}_j}\right]
\nonumber\\
{\bf {\cal W}}^+_{ij}&\equiv&
-\left(\frac{{\bf r}_{ij}}{2}\times \left[{\bf \omega}_i
+{\bf \omega}_j\right]\right)
+\frac{k_BT}{I}
\left( \frac{{\bf r}_{ij}}{2}\times 
\left[\frac{\partial}{\partial{\bf \omega}_i}
+\frac{\partial}{\partial{\bf \omega}_j}\right]\right)
\label{rate2b}
\end{eqnarray}
to be compared with (\ref{sum6}).

\subsection{Balance equations}

The conserved density fields are expected to behave hydrodynamically.  The conserved
density fields are the mass density $\rho({\bf r},t)= m n({\bf r},t)$,
where $n({\bf r},t)$ is the number density field; the momentum density
$\rho({\bf r},t){\bf u}({\bf r},t)$, where ${\bf u}({\bf r},t)$ is the
velocity field; and the total angular momentum density field ${\bf
J}({\bf r},t)= {\bf L}({\bf r},t)+{\bf S}({\bf r},t)$ where ${\bf
L}({\bf r},t)={\bf r}\times \rho({\bf r},t){\bf u}({\bf r},t)$ is the
macroscopic angular momentum density and ${\bf S}({\bf r},t)=In({\bf
r},t){\bf \Omega}({\bf r},t)$ is the intrinsic angular momentum
density or spin density. Here ${\bf \Omega}({\bf r},t)$ is the angular
velocity field.  The number density, the velocity and angular velocity
fields are defined by

\begin{eqnarray}
n({\bf r},t) &=& \langle \sum_i \delta({\bf r}-{\bf r}_i)\rangle
\nonumber\\
n({\bf r},t){\bf u}({\bf r},t)
&=& \langle \sum_i {\bf v}_i\delta({\bf r}-{\bf r}_i)\rangle
\nonumber\\
n({\bf r},t){\bf \Omega}({\bf r},t)
&=& \langle \sum_i {\bf \omega}_i\delta({\bf r}-{\bf r}_i)\rangle
\label{cons}
\end{eqnarray}
By applying (\ref{rate1}) to the mass and momentum densities
(\ref{cons}) we obtain the set of balance equations
\begin{eqnarray}
\partial_t\rho&=&-\nabla\rho{\bf u}
\nonumber\\
\partial_t\rho{\bf u}&=&-\nabla\left[\rho{\bf u}{\bf u}
+{\bf \Pi}\right]
\nonumber \\
\label{balance}
\end{eqnarray}
where the total stress tensor ${\bf \Pi}={\bf \Pi}^K+{\bf \Pi}^C+{\bf
\Pi}^D$ and the kinetic, conservative and dissipative contributions to the stress
tensor are
\begin{eqnarray}
{\bf \Pi}^K&=&\langle m \sum_i ({\bf v}_i-{\bf u}({\bf r},t))
({\bf v}_i-{\bf u}({\bf r},t))\delta({\bf r}-{\bf r}_i)\rangle
\nonumber\\
{\bf \Pi}^C&=&\langle\frac{1}{2}\sum_{i,j\neq i}{\bf F}^C_{ij}{\bf r}_{ij}
\int_0^1 d\lambda\delta({\bf r}-{\bf r}_{i}-\lambda{\bf r}_{ij})\rangle
\nonumber\\
{\bf \Pi}^D&=&-\gamma m
\langle\frac{1}{2}\sum_{i,j\neq i}{\bf r}_{ij}{\bf T}_{ij}\!\cdot\!{\bf g}_{ij}
\int_0^1 d\lambda\delta({\bf r}-{\bf r}_{i}-\lambda{\bf r}_{ij})\rangle
\label{stress}
\end{eqnarray}
Here, ${\bf g}_{ij}={\bf v}_{ij}+{\bf
r}_{ij}\times[\omega_i+\omega_j]/2$ is the relative velocity at the
``surface of contact'' of two identical spheres separated a distance
$r_{ij}$. In deriving these equations we have used the identity
\begin{eqnarray}
\delta({\bf r}-{\bf r}_i)-\delta({\bf r}-{\bf r}_j)
&=&\int_0^1d\lambda\frac{d}{d\lambda}
\delta({\bf r}-{\bf r}_i+\lambda {\bf r}_{ij})
\nonumber\\
&=&-\nabla\!\cdot\!{\bf r}_{ij}\int_0^1d\lambda
\delta({\bf r}-{\bf r}_i+\lambda {\bf r}_{ij})
\label{taylor}
\end{eqnarray}
We note that the kinetic and conservative parts to the stress tensor
are symmetric tensors but the dissipative part is not and
therefore we must be careful with the ordering of the indices. In
Cartesian components we understand the momentum balance equation
(\ref{balance}) as follows (summation over repeated indices is implied)
\begin{equation}
\partial_t \rho {\bf u}^\nu =
\partial_\mu \left[\rho{\bf u}^\mu{\bf u}^\nu
+{\bf \Pi}_{\mu\nu}\right]
\label{indor}
\end{equation}
and the dissipative stress tensor has the form
\begin{equation}
{\bf \Pi}^D_{\mu\nu}=-\gamma m
\langle\frac{1}{2}\sum_{i,j\neq i}
{\bf r}^{\mu}_{ij}{\bf T}^{\nu\alpha}_{ij}{\bf g}^{\alpha}_{ij}
\int_0^1 d\lambda
\delta({\bf r}-{\bf r}_{i}-\lambda{\bf r}_{ij})\rangle
\label{indor2}
\end{equation}

Concerning the angular velocity field, by using again
Eqn. (\ref{rate1}) on the definition (\ref{cons}) we obtain

\begin{equation}
\partial_t n{\bf \Omega}=
-\nabla\langle\sum_i{\bf v}_i{\bf \omega}_i
\delta({\bf r}-{\bf r}_i)\rangle
+\frac{m}{I}\gamma\langle\sum_{i,j\neq i}
\left(\frac{{\bf r}_{ij}}{2}\times
{\bf T}_{ij}\!\cdot\!{\bf g}_{ij}\right)\delta({\bf r}-{\bf r}_i)\rangle
\label{no}
\end{equation}
 Note that the rate of change of the spin ${\bf S}=In{\bf
\Omega}$ cannot be expressed entirely as the gradient of a flux. This
is a reflection that the intrinsic angular momentum ${\bf S}$ is not a
conserved quantity. In the same way, the macroscopic angular momentum
${\bf L}={\bf r}\times \rho{\bf u}$ is not conserved either, as can
be appreciated by taking the cross product of the momentum balance
equation (\ref{balance}) with the position vector ${\bf r}$, this is

\begin{eqnarray}
\partial_t {\bf L}&=&-{\bf r}\times \nabla (\rho{\bf u}{\bf u}+{\bf \Pi})
\nonumber\\
&=&- \nabla ({\bf L}{\bf u}+{\bf r}\times{\bf \Pi})+2{\bf \Pi}^A
\label{L}
\end{eqnarray}
where ${\bf \Pi}^A$ is the antisymmetric part of the stress tensor
(expressed here as an axial vector, this is ${\bf
\Pi}^{A\alpha}=\frac{1}{2}\epsilon^{\alpha\mu\nu}{\bf \Pi}^{\mu\nu}$
where $\epsilon^{\alpha\mu\nu}$ is the Levi-Civita symbol). If the
stress tensor is symmetric (i.e. its antisymmetric part is zero), the
macroscopic angular momentum is conserved. In the fluid particle model
the non-central nature of the forces implies that the antisymmetric
part of the stress tensor is not zero. Actually, it is given by (as an
axial vector)

\begin{equation}
2{\bf \Pi}^A = -\gamma m 
\langle \sum_{i,j\neq i}\left(\frac{{\bf r}_{ij}}{2}
\times {\bf T}_{ij}\!\cdot\!{\bf g}_{ij}\right)
\int_0^1d\lambda
\delta({\bf r}-{\bf r}_i+\lambda {\bf r}_{ij})\rangle
\label{pia}
\end{equation}

If we add the last term of (\ref{no}) with the last term of (\ref{L})
which is (\ref{pia}), we obtain

\begin{eqnarray}
2{\bf \Pi}^A +
\gamma m\langle \sum_{i,j\neq i}\left(\frac{{\bf r}_{ij}}{2}
\times {\bf T}_{ij}\!\cdot\!{\bf g}_{ij}\right)
\delta ({\bf r}-{\bf r}_i)\rangle
&=&
\gamma m\langle \sum_{i,j\neq i}\left(\frac{{\bf r}_{ij}}{2}
\times {\bf T}_{ij}\!\cdot\!{\bf g}_{ij}\right)
\left[\delta ({\bf r}-{\bf r}_i)-
\int_0^1d\lambda
\delta({\bf r}-{\bf r}_i+\lambda {\bf r}_{ij})\right]\rangle
\nonumber\\
&=&
\gamma m \nabla\langle \sum_{i,j\neq i}\left(\frac{{\bf r}_{ij}}{2}
\times {\bf T}_{ij}\!\cdot\!{\bf g}_{ij}\right){\bf r}_{ij}
\int_0^1d\lambda\int_0^1d\lambda'
\delta({\bf r}-{\bf r}_i+\lambda \lambda'{\bf r}_{ij})\rangle
\label{sour}
\end{eqnarray}
where we have used again (\ref{taylor}). Therefore, the {\em total}
angular momentum density ${\bf J}= {\bf L}+{\bf S}$ satisfies a
balance equation

\begin{equation}
\partial_t {\bf J}=-\nabla\left[{\bf J}{\bf v}+{\bf r}\times {\bf \Pi}
+ {\bf \Phi}\right]
\label{bj}
\end{equation}
where
\begin{equation}
{\bf \Phi}=
\gamma m \langle\sum_{i,j\neq i}
\left(\frac{{\bf r}_{ij}}{2}\times
{\bf T}_{ij}\!\cdot\!{\bf g}_{ij}\right){\bf r}_{ij}
\int_0^1d\lambda\int_0^1d\lambda'\delta({\bf r}-{\bf r}_i
+\lambda \lambda'{\bf r}_{ij})\rangle
\label{bjb}
\end{equation}

\subsection{Balance equations in terms of distribution functions}

It is convenient to express the quantities appearing in the balance
equations in terms of the single particle and pair distribution
functions, defined as

\begin{eqnarray}
f({\bf x},t) &=&f({\bf r},{\bf v},{\bf \omega},t) 
= \langle \sum_i \delta({\bf x}-{\bf x}_i)\rangle
\nonumber\\
f^{(2)}({\bf x},{\bf x'},t) &=&
 \langle \sum_{i,j\ne i} \delta({\bf x}-{\bf x}_i)
 \delta({\bf x'}-{\bf x}_j)\rangle
\label{fs}
\end{eqnarray}
The number density, the velocity and angular velocities
in (\ref{cons}) are the first moments of the single particle
distribution function,
\begin{eqnarray}
n({\bf r},t) &=& 
\int d{\bf v}d{\bf \omega}f({\bf r},{\bf v},{\bf \omega},t) 
\nonumber\\
n({\bf r},t){\bf u}({\bf r},t)
&=& \int d{\bf v}d{\bf \omega}{\bf v}f({\bf r},{\bf v},{\bf \omega},t) 
\nonumber\\
n({\bf r},t){\bf \Omega}({\bf r},t)
&=& \int d{\bf v}d{\bf \omega}{\bf \omega}f({\bf r},{\bf v},{\bf \omega},t) 
\label{consb}
\end{eqnarray}
Next, by using that for an arbitrary function $G$
\begin{eqnarray}
\langle\sum_{i,j\neq i} 
G({\bf r}_{ij},{\bf v}_i,{\bf v}_j,{\bf \omega}_i,{\bf \omega}_j)
\delta({\bf r}-{\bf r}_i+\lambda{\bf r}_{ij})
\rangle
&=&
\int d{\bf v}d{\bf v}'d{\bf \omega}d{\bf \omega}'
d{\bf R}
G({\bf R},{\bf v},{\bf v}',{\bf \omega},{\bf \omega}')
\nonumber\\
&\times&f^{(2)}({\bf r}+\lambda{\bf R},
{\bf v},{\bf \omega},{\bf r}+(\lambda-1){\bf R},{\bf v}',{\bf \omega}')
\label{distrib}
\end{eqnarray}
which, for $\lambda=0$ becomes
\begin{eqnarray}
\langle\sum_{i,j\neq i} 
G({\bf r}_{ij},{\bf v}_i,{\bf v}_j,{\bf \omega}_i,{\bf \omega}_j)
\delta({\bf r}-{\bf r}_i)
\rangle
&=&
\int d{\bf v}d{\bf v}'d{\bf \omega}d{\bf \omega}'
d{\bf R}
G({\bf R},{\bf v},{\bf v}',{\bf \omega},{\bf \omega}')
\nonumber\\
&\times&
f^{(2)}({\bf r},{\bf v},{\bf \omega},{\bf r}-{\bf R},{\bf v}',{\bf \omega}')
\label{dist2}
\end{eqnarray}
we can write the different contributions (\ref{stress}) to the stress
tensor in terms of the distribution functions

\begin{eqnarray}
{\bf \Pi}^K&=&\int d{\bf v} d{\bf \omega}
 m({\bf v}-{\bf u})({\bf v}-{\bf u})
f({\bf r},{\bf v},{\bf \omega},t)
\nonumber\\
{\bf \Pi}^C&=&
\int d{\bf v} d{\bf \omega}d{\bf v}' d{\bf \omega}'\int d{\bf R}
\frac{{\bf R}}{2}{\bf F}^C({\bf R})
{\overline f}^{(2)}({\bf r},{\bf v},{\bf \omega},
{\bf r}',{\bf v}',{\bf \omega}')
\nonumber\\
{\bf \Pi}^D&=&
-\gamma m
\int d{\bf v} d{\bf \omega}d{\bf v}' d{\bf \omega}'\int d{\bf R}
\frac{{\bf R}}{2}{\bf T}({\bf R})\!\cdot\!{\bf g}
{\overline f}^{(2)}({\bf r},{\bf v},{\bf \omega},
{\bf r}',{\bf v}',{\bf \omega}')
\label{stressdist}
\end{eqnarray}
where ${\bf g}\equiv\left[{\bf v}-{\bf v}' +\frac{{\bf
R}}{2}\times[{\bf \omega}+{\bf \omega}']\right]$. We have introduced
in these expressions the spatially averaged pair distribution function

\begin{equation}
{\overline f}^{(2)}({\bf r},{\bf v},{\bf \omega},
{\bf r}',{\bf v}',{\bf \omega}')=
\int_0^1d\lambda f^{(2)}({\bf r}+\lambda{\bf R},
{\bf v},{\bf \omega},{\bf r}+(\lambda-1){\bf R},{\bf v}',{\bf \omega}')
\label{spa2}
\end{equation}

In terms of the distribution functions the terms of the right hand side of
(\ref{no}) can be written as
\begin{eqnarray}
\langle \sum_{i,j\neq i}{\bf v}_i{\bf \omega}_i 
\delta ({\bf r}-{\bf r}_i)\rangle
&=&
\int d{\bf v} d{\bf \omega} {\bf v}{\bf \omega}
f({\bf r},{\bf v},{\bf \omega},t)
\nonumber\\
&=&
\int d{\bf v} d{\bf \omega} 
({\bf v}-{\bf u})({\bf \omega}-{\bf \Omega})
f({\bf r},{\bf v},{\bf \omega},t)-n{\bf \Omega}{\bf u}
\nonumber\\
&+&{\bf u}\int d{\bf v} d{\bf \omega} {\bf \omega}
f({\bf r},{\bf v},{\bf \omega},t)
+{\bf \Omega}\int d{\bf v} d{\bf \omega} {\bf v}
f({\bf r},{\bf v},{\bf \omega},t)
\nonumber\\
\langle \sum_{i,j\neq i}\left(\frac{{\bf r}_{ij}}{2}
\times {\bf T}_{ij}\!\cdot\!{\bf g}_{ij}\right)
\delta ({\bf r}-{\bf r}_i)\rangle
&=&
\int d{\bf v} d{\bf \omega}d{\bf v}' d{\bf \omega}'\int d{\bf R}
\left(\frac{{\bf R}}{2}\times{\bf T}({\bf R})
\!\cdot\!{\bf g}\right)
\nonumber\\
&\times&
 f^{(2)}({\bf r},{\bf v},{\bf \omega}, 
{\bf r}-{\bf R},{\bf v}',{\bf \omega}')
\label{cont2}
\end{eqnarray}
Finally,

\begin{equation}  
{\bf \Phi}=
\gamma m \int d{\bf v} d{\bf \omega}d{\bf v}' d{\bf \omega}'\int d{\bf R}
\left(\frac{{\bf R}}{2}\times{\bf T}({\bf R})\!\cdot\!{\bf g}\right)
{\bf R}
{\overline {\overline f}}^{(2)}({\bf r},{\bf v},{\bf \omega},{\bf r}',{\bf v}',{\bf \omega}')
\label{stressdistb}
\end{equation}
where
\begin{equation}
{\overline {\overline f}}^{(2)}({\bf r},{\bf v},{\bf \omega},{\bf r}',{\bf v}',{\bf \omega}')
=\int_0^1d\lambda\int_0^1d\lambda'
f^{(2)}({\bf r}+\lambda\lambda'{\bf R},{\bf v},{\bf \omega},{\bf r}
+(\lambda\lambda'-1){\bf R},{\bf v}',{\bf \omega}')
\label{oof}
\end{equation}
\subsection{Fokker-Planck-Boltzmann equation}
The Fokker-Planck-Boltzmann equation (FPBE) is an approximate kinetic
equation for the single particle distribution function $f({\bf x},t)$.
The FPBE is obtained by applying the general rate of change equation
(\ref{rate1}) to $f({\bf x},t)$. After some algebra one arrives at

\begin{eqnarray}
\partial_t f+{\bf v}\!\cdot\!\nabla f 
&=&
\int d{\bf R}d{\bf v}'d{\bf \omega}'
{\bf \partial}\!\cdot\!\left[ {\bf F}^C({\bf R})+\gamma {\bf T}({\bf R})\!\cdot\!
{\bf g}\right]
f^{(2)}({\bf r},{\bf v},{\bf \omega},{\bf r}-{\bf R},{\bf v}',{\bf \omega}')
\nonumber\\
&+&
\frac{k_BT}{m}\gamma\int d{\bf R}d{\bf v}'d{\bf \omega}'
{\bf \partial}\!\cdot\!{\bf T}({\bf R})\!\cdot\!{\bf \partial}
f^{(2)}({\bf r},{\bf v},{\bf \omega},{\bf r}-{\bf R},{\bf v}',{\bf \omega}')
\label{bbgky1}
\end{eqnarray}
where we have defined the operator

\begin{equation}
\partial \equiv \frac{\partial}{\partial{\bf v}}+
\frac{m}{I}\left(\frac{{\bf R}}{2}\times\frac{\partial}{\partial \omega}\right)
\label{partial}
\end{equation}
In obtaining (\ref{bbgky1}), we have inserted at some point the identity
\begin{equation}
1=\int d{\bf R}d{\bf v}'d{\bf \omega}'
\delta({\bf r}-{\bf R}-{\bf r}_j)
\delta({\bf v}'-{\bf v}_j)
\delta({\bf \omega}'-{\bf \omega}_j)
\label{i}
\end{equation}
Equation (\ref{bbgky1}) is not a closed equation for $f({\bf r},{\bf
v},{\bf \omega},t)$ because the pair function $f^{(2)}({\bf x},{\bf
x}',t)$ appears. Nevertheless it can be closed approximately by using
the molecular chaos assumption. In what follows we will assume that
the friction $\gamma$ is so large to allow for a neglection of the
conservative forces ${\bf F}^C$ \cite{mar97}. This simplifies
considerably the calculations in the next section. The molecular chaos
assumption in the absence of conservative forces becomes

\begin{equation}
f^{(2)}({\bf x},{\bf x}',t)\approx f({\bf x},t)f({\bf x}',t)
\label{chaos}
\end{equation}

The final closed Fokker-Planck-Boltzmann equation for
the distribution function is, after using the molecular chaos
assumption (\ref{chaos}) 

\begin{equation}
\partial_t f+{\bf v}\!\cdot\!\nabla f 
=I[f]=
\gamma\int d{\bf R}d{\bf v}'d{\bf \omega}'
f({\bf r}-{\bf R},{\bf v}',{\bf \omega}')
{\bf \partial} \!\cdot\!{\bf T}({\bf R})\!\cdot\!
\left[{\bf g}+\frac{k_BT}{m}{\bf \partial}\right]
f({\bf r},{\bf v},{\bf \omega})
\label{fpbe}
\end{equation}
which is an integro-differential non-linear equation.

\subsection{Chapman-Enskog solution of the FPBE}
Our aim is to solve the nonlinear FPBE (\ref{fpbe}) by using the
perturbative method of Chapman and Enskog. The method is valid for
situations in which the macroscopic conserved fields are slowly
varying in typical molecular length scales. In these situations, the
distribution function decays in a very short kinetic time (short
compared to typical times of evolution of the conserved field) towards
the so called {\em normal} solution where the distribution function
$f({\bf v},{\bf \omega}|a({\bf r},t))$ depends on space and time only
through the first few moments $a({\bf r},t)$ \cite{mar97}.  During
this last hydrodynamic stage, the solution can be obtained
perturbatively as an expansion in gradients, this is $f({\bf v},{\bf
\omega}|{\bf u},{\bf \Omega})= f_0+ f_1+{\cal O}(\nabla^2)$ where
$f_0$ is of zeroth order in gradients and $f_1$ is of first order in
gradients.  By substitution of this expansion into the FPBE
(\ref{fpbe}) one obtains

\begin{equation}
\partial_t f_0 +\partial_t f_1+ {\bf v}\!\cdot\!\nabla f_0
=
I[f_0]+ \left(\frac{d I}{df}\right)_{f_0}f_1+{\cal O}(\nabla^2)
\label{exp1}
\end{equation}
By analogy with the conventional kinetic theory and also with the
kinetic theory for DPD in Ref. \cite{mar97}, we expect that the lowest
order contribution $f_0$ is given by the {\em local equilibrium} form
for the distribution function. In the presence of spin variables it
takes the form

\begin{equation}
f_{0}({\bf r},{\bf v},{\bf \omega},t)
=
n({\bf r},t)\left(\frac{m}{2\pi k_BT}\right)^{D/2}
\exp\left\{-\frac{m}{2k_BT}({\bf v}-{\bf u})^2\right\}
\left(\frac{I}{2\pi k_BT}\right)^{D/2}
\exp\left\{-\frac{I}{2k_BT}({\bf \omega}-{\bf \Omega})^2\right\}
\label{f0}
\end{equation}
This local equilibrium distribution provides the correct averages for
the first moments of $f({\bf r},{\bf v},{\bf \omega},t)$, this is,
\begin{eqnarray}
n({\bf r},t) &=& \int d{\bf v}d{\bf \omega}f_{0}({\bf r},{\bf v},{\bf \omega},t) 
\nonumber\\
n({\bf r},t){\bf u}({\bf r},t)
&=& \int d{\bf v}d{\bf \omega}{\bf v}f_{0}({\bf r},{\bf v},{\bf \omega},t) 
\nonumber\\
n({\bf r},t){\bf \Omega}({\bf r},t)
&=& 
\int d{\bf v}d{\bf \omega}{\bf \omega}f_{0}({\bf r},{\bf v},{\bf \omega},t) 
\label{cons1}
\end{eqnarray}
This, in turn, implies that

\begin{eqnarray}
\int d{\bf v}d{\bf \omega}f_{1}({\bf r},{\bf v},{\bf \omega},t)  
&=& {\cal O}(\nabla^2)
\nonumber\\
\int d{\bf v}d{\bf \omega}{\bf v}f_{1}({\bf r},{\bf v},{\bf \omega},t) 
&=& {\cal O}(\nabla^2)
\nonumber\\
\int d{\bf v}d{\bf \omega}{\bf \omega}f_{1}({\bf r},{\bf v},{\bf \omega},t) 
&=& {\cal O}(\nabla^2)
\label{consf1}
\end{eqnarray}

The procedure is now a bit different from the Chapman-Enskog method in
Ref. \cite{mar97}, because the inclusion of the spin variables
produces new terms with different orders in gradients. We write the
equation (\ref{exp1}) as follows
\begin{equation}
\left(\frac{d I}{df}\right)_{f_0}f_1-\partial_t f_1=
\partial_t f_0 + {\bf v}\!\cdot\!\nabla f_0-I[f_0] 
\label{exp1bis}
\end{equation}
where we have neglected terms that are quadratic in gradients. We will
check in the following that both sides of this equation are
explicitly of first order in gradients. This linear equation
(\ref{exp1bis}) will be solved for $f_1$ and therefore we will obtain
the solution of the FPBE (\ref{fpbe}) as $f_{0}+ f_1$, up to terms of order
$\nabla^2$.

We now consider each term of (\ref{exp1bis}) separately. The temporal
and spatial derivatives of $f_0$ can be computed to first order in
gradients with the use of the balance equations (\ref{balance}) and (\ref{no}).
Only terms of order $\nabla$ are to be retained, which amounts to use the
balance equations with the averages of the quantities appearing in
them evaluated with the local equilibrium ensemble. Therefore, we need
to compute the local equilibrium average of the stress tensor ${\bf
\Pi}$ in the momentum balance equation, and the local equilibrium
average of the two contribution in (\ref{cont2}) to the equation for
the angular velocity field. After using the molecular chaos assumption
one easily obtains the following results

\begin{eqnarray}
{\bf \Pi}^{K0}_{\mu\nu}&=&n k_BT\delta_{\mu\nu}
\nonumber\\
{\bf \Pi}^{D0}_{\mu\nu}&=&-\gamma m n^2\frac{1}{2}\left[
A_2\left[\partial_\nu{\bf u}^\mu+
\epsilon^{\mu\nu\sigma}{\bf \Omega}^\sigma\right]+
B_2\left[\nabla\!\cdot\!{\bf u}\delta_{\mu\nu}+
\partial_\nu{\bf u}^\mu+\partial_\mu{\bf u}^\nu\right]\right]
+ {\cal O}(\nabla^2)
\label{stress0}
\end{eqnarray}
where we have defined
\begin{eqnarray}
A_2&\equiv& \frac{1}{D}\int d{\bf R} R^2 A(R)
\nonumber\\
B_2&\equiv& \frac{1}{D(D+2)}\int d{\bf R} R^2 B(R)
\label{oA}
\end{eqnarray}
The first contribution ${\bf \Pi}^{K0}$ produces an isotropic pressure
term. Consistently with our assumption that the conservative forces
are negligible this pressure is given by the ideal gas expression. The
second contribution ${\bf \Pi}^{D0}$ contains terms of first order in
gradients. We arrange a bit this contribution by introducing the
velocity gradient tensor $(\nabla {\bf u})_{\nu\mu}=\partial_\nu{\bf
u}^\mu$ and its traceless symmetric and antisymmetric parts

\begin{eqnarray}
\overline{\nabla {\bf u}}^S&\equiv&\frac{1}{2}\left[
\nabla {\bf u}+\nabla {\bf u}^T\right]-\frac{1}{D}\nabla\!\cdot\!{\bf u 1}
\nonumber\\
\nabla {\bf u}^A&\equiv&\frac{1}{2}\left[\nabla {\bf u}-\nabla {\bf u}^T\right]
\label{velgrad}
\end{eqnarray}
We have

\begin{equation}
{\bf \Pi}^{D0}=-\gamma m n^2\frac{1}{2}\left[
A_2(\nabla {\bf u}^A+{\bf \Omega}) 
+(A_2+2B_2)\overline{\nabla {\bf u}}^S
+\left[A_2+(D+2)B_2\right]\frac{1}{D}
\nabla\!\cdot\!{\bf u 1}\right]
\label{pito}
\end{equation}
The antisymmetric part of the total stress tensor in the local equilibrium
approximation to first order in gradients is given by (as an axial vector)

\begin{equation}
{\bf \Pi}^{A0}=\gamma m n^2 
\frac{A_2}{2}
\left(\frac{1}{2}\nabla\times {\bf u}-{\bf \Omega}\right)
\label{apito}
\end{equation}
In a similar way one computes the quantities in (\ref{cont2}) that
appear in the balance equation for the spin (\ref{no}). In particular,
the last term in (\ref{cont2}) is also given by $-2{\bf \Pi}^A$ in the
local equilibrium approximation to first order gradients.

Substitution of the local equilibrium expressions for the stress tensor
into the balance equations produce the Euler equations,

\begin{eqnarray}
\partial_t n&=&-\nabla n{\bf u}
\nonumber\\
\partial_t{\bf u}&=&-({\bf u}\!\cdot\!\nabla){\bf u}
-\frac{k_BT}{m} \frac{1}{n}\nabla n
+\frac{1}{n}\nabla\times \gamma\frac{A_2}{2}n^2{\bf \Omega}
\nonumber \\
\partial_t {\bf \Omega}&=&-({\bf u}\!\cdot\!\nabla){\bf \Omega}
+\gamma\frac{m}{I}A_2n
\left[\frac{1}{2}\nabla\times {\bf u}
-{\bf \Omega}\right]
\label{consle}
\end{eqnarray}
We have neglected a term of first order in gradients which produces a
term of order $\nabla^2$ in the momentum balance equation. We note
that the time derivative of the angular velocity contains a term which
is of {\em zeroth} order in gradients (the ${\bf \Omega}$ term in the 
last equation).

With the help of the Euler equations and the chain rule, we can now compute
the time and space derivatives of the local equilibrium distribution,
to first order in gradient. The result is

\begin{eqnarray}
\partial_t f_0+ {\bf v}\!\cdot\!\nabla f_0
&=&f_0\left[-
\nabla\!\cdot\!{\bf u}
+\frac{m}{k_BT}
({\bf v}-{\bf u})({\bf v}-{\bf u}):\nabla{\bf u}
+\frac{I}{k_BT}
({\bf \omega}-{\bf \Omega})({\bf v}-{\bf u}):\nabla{\bf \Omega}
\right.
\nonumber\\
&+&\left.\gamma n A_2\frac{m}{k_BT}\frac{1}{2n}
({\bf v}-{\bf u})\!\cdot\!
\nabla\times n^2{\bf \Omega}
+\gamma A_2n\frac{m}{k_BT}({\bf \omega}-{\bf \Omega})
\!\cdot\!\left[\frac{1}{2}\nabla\times {\bf u}-{\bf \Omega}\right]
\right]
\label{f0a}
\end{eqnarray}
where the double dot $:$ denotes double contraction.

The next step is the calculation of $I[f_0]$. To first order in gradients it
is given by

\begin{equation}
I[f_0]=\gamma nA_2 \frac{m}{k_BT}f_0
\left[
\frac{1}{2n}({\bf v}-{\bf u})\!\cdot\!\nabla\times n^2{\bf \Omega}
+({\bf \omega}-{\bf \Omega})\!\cdot\!
\left[\frac{1}{2}\nabla\times {\bf u}-{\bf \Omega}\right]\right]
\label{if0}
\end{equation}
Therefore, after some happy cancelations the right hand side of
(\ref{exp1bis}) has the simple form

\begin{equation}
\partial_t f_0+ {\bf v}\!\cdot\!\nabla f_0-I[f_0]=
f_0\left[-\nabla\!\cdot\!{\bf u}
+\frac{m}{k_BT}({\bf v}-{\bf u})({\bf v}-{\bf u}):\nabla{\bf u}
+\frac{I}{k_BT}({\bf \omega}-{\bf \Omega})({\bf v}-{\bf u}):\nabla{\bf \Omega}
\right]
\label{f0-i0}
\end{equation}
which contains only terms of first order in gradients. 

Next, we consider the term $\partial_tf_1$ in (\ref{exp1bis}). We note that it is of
first order in gradients due to the term of zeroth order in the Euler
equation for the angular velocity, this is
\begin{eqnarray}
\partial_t f_1 &=& \frac{\partial f_1}{\partial {\bf \Omega}}\partial_t{\bf \Omega} +{\cal O}(\nabla^2)
\nonumber\\
&=& -\frac{m}{I}\gamma A_2 n {\bf \Omega}\frac{\partial f_1}{\partial {\bf \Omega}}
 +{\cal O}(\nabla^2)
\nonumber\\
&=& \frac{m}{I}\gamma A_2 n {\bf \Omega}\frac{\partial f_1}{\partial {\bf \omega}}
 +{\cal O}(\nabla^2)
\end{eqnarray}
where we have assumed that the dependence of $f_1$ on ${\bf \Omega}$ appears in
the combination ${\bf \omega} -{\bf \Omega}$. This assumption will be confirmed
{\em a posteriori}.

The linearization of the functional $I[f]$ might be easier to perform
by expanding $I[f_0+f_1]$ to first order in gradients (at some point
one uses Eqns. (\ref{consf1})). The final result for
the left hand side of (\ref{exp1bis}) is

\begin{equation}
I[f_0+f_1]-I[f_0]-\partial_t f_1 
=\gamma n\left[[A_0+B_0]{\cal L}^T
+\frac{m}{I}\frac{A_2}{2}{\cal L}^R\right]f_1
+{\cal O}(\nabla^2)
\label{lhs}
\end{equation}
where the operators are given by

\begin{eqnarray}
{\cal L}^T&=&
\frac{\partial}{\partial {\bf v}}\!\cdot\!
\left[{\bf v}-{\bf u}
+\frac{k_BT}{m}\frac{\partial}{\partial {\bf v}}\right]
\nonumber\\
{\cal L}^R&=&
\frac{\partial}{\partial {\bf \omega}}\!\cdot\!
\left[{\bf \omega}-{\bf \Omega}
+\frac{k_BT}{I}\frac{\partial}{\partial {\bf \omega}}\right]
\label{oprt}
\end{eqnarray}
and the constants $A_0,B_0$ are given by

\begin{eqnarray}
A_0&\equiv& \int d{\bf R} A(R)
\nonumber\\
B_0&\equiv &\frac{1}{D}\int d{\bf R}  B(R)
\label{const}
\end{eqnarray}

Eqn. (\ref{exp1bis}) can be written in compact form as

\begin{equation}
{\cal L}f_1=
f_0\left[-\nabla\!\cdot\!{\bf u}
+\frac{m}{k_BT}{\bf V}{\bf V}:\nabla{\bf u}
+\frac{I}{k_BT}{\bf O}{\bf V}:\nabla{\bf \Omega}
\right]
\label{eqcom}
\end{equation}
where the operator has the form

\begin{equation}
{\cal L}= \gamma n \left[[A_0+B_0]{\cal L}^T
+\frac{m}{I}\frac{A_2}{2}{\cal L}^R\right]
\label{opeeqcom}
\end{equation}
and the peculiar velocities are ${\bf V}={\bf v}-{\bf u},
{\bf O}={\bf \omega}-{\bf \Omega}$. 

Equation (\ref{eqcom}) is an inhomogeneous second order partial
differential equation. In order to obtain a special solution of the
inhomogeneous equations (\ref{eqcom}), we introduce the following
tensors

\begin{eqnarray}
{\bf J}_{\mu\nu}&=&\frac{m}{k_BT}
\left\{{\bf V}_\mu{\bf V}_\nu
-\frac{1}{D}V^2\delta_{\mu\nu}\right\}
\nonumber\\
{\cal J}&=&\frac{m V^2}{Dk_BT}-1
\nonumber\\
{\cal  T}_{\mu\nu}&=&\frac{I}{k_BT}
{\bf O}_\mu{\bf V}_\nu
\label{defin}
\end{eqnarray}
With these quantities we write (\ref{eqcom}) in the form

\begin{equation}
{\cal L}f_1=
f_0\left[{\bf J}:\overline{\nabla {\bf u}}^S
+{\cal J}\nabla\!\cdot\!{\bf u}{\bf 1}
+{\cal  T}:\nabla{\bf \Omega}\right]
\label{eqcom2}
\end{equation}
The quantities (\ref{defin}) satisfy

\begin{eqnarray}
{\cal L}^Tf_0{\bf J}_{\mu\nu}&=&-2f_0{\bf J}_{\mu\nu}
\nonumber\\
{\cal L}^Rf_0{\bf J}_{\mu\nu}&=&0
\nonumber\\
{\cal L}^Tf_0{\cal J}&=&-2f_0{\cal J}
\nonumber\\
{\cal L}^Rf_0{\cal J}&=&0
\nonumber\\
{\cal L}^Tf_0{\cal T}_{\mu\nu}&=&f_0{\cal T}_{\mu\nu}
\nonumber\\
{\cal L}^Rf_0{\cal T}_{\mu\nu}&=&f_0{\cal T}_{\mu\nu}
\label{eigen}
\end{eqnarray}
and therefore, a special solution of (\ref{eqcom}) is given by
\begin{equation}
f_1 = -f_0\left[\frac{1}{2\gamma n(A_0+B_0)}
\left[{\bf J}:\overline{\nabla {\bf u}}^S+{\cal J}\nabla\!\cdot\!{\bf
u}{\bf 1}\right] -\frac{1}{\gamma n(A_0+B_0+\frac{m}{I}\frac{A_2}{2})} {\cal T}:\nabla{\bf
\Omega}\right]
\label{f1sol}
\end{equation}
as can be checked by substitution.

Now it remains to obtain a general solution of the homogeneous
equation ${\cal L}f_1=0$. The solution of this homogeneous equation is
an arbitrary linear combination of $f_0a$, where $a$ are the
collisional invariants $a=\{1,{\bf v}-{\bf u},{\bf \omega}-{\bf
\Omega}\}$. Nevertheless, the combination of (\ref{consb}) and
(\ref{cons1}) imposes that the coefficients of the linear combination
are zero.

\subsection{Transport coefficients}

The phenomenological theory of viscous flow of an isotropic fluid
\cite{gro62} relates the trace ${\rm tr}\left[{\bf \Pi}\right]$, the
traceless symmetric ${\overline {\bf \Pi}}^S$ and antisymmetric
${\bf\Pi}^A$ parts of the stress tensor ${\bf \Pi}$ with the linear
velocity gradients and angular velocity in the following way

\begin{eqnarray}
\frac{1}{D}{\rm tr}\left[{\bf \Pi}\right]&=&-\zeta \nabla\!\cdot\!{\bf u} + p
\nonumber\\
{\overline {\bf \Pi}}^S&=&-2\eta\overline{\nabla {\bf u}}^S
\nonumber\\
{\bf \Pi}^A&=&-2\eta_R\left[\frac{1}{2}\nabla\times{\bf u}-{\bf \Omega}
\right]
\label{phen}
\end{eqnarray}
where the antisymmetric part is written as an axial vector. Here $p$
is the isotropic hydrostatic pressure.  The coefficients are the bulk
viscosity $\zeta$, the shear viscosity $\eta$ and the rotational
viscosity $\eta_R$.

We now compute the stress tensor (\ref{stressdist}) using the
molecular chaos assumption (\ref{chaos}) for the pair distribution
function and the approximate solution $f_0+f_1$ for the single
particle distribution function. This will produce ${\bf \Pi}={\bf
\Pi}_0+{\bf \Pi}_1$ where the local equilibrium contribution ${\bf
\Pi}_0$ has been already computed in (\ref{stress0}). Regarding the
term ${\bf \Pi}_1$ computed with $f_1$ one observes that the only
contribution which is of first order in gradients is ${\bf \Pi}^K_1$
which is computed along similar lines to Ref. \cite{mar97}. The final
result is

\begin{equation}
{\bf \Pi}^K=nk_BT{\bf 1}
-\frac{k_BT}{\gamma[A_0+B_0]}
\overline{\nabla {\bf u}}^S
-\frac{k_BT}{D\gamma[A_0+B_0]}
\nabla\!\cdot\!{\bf u 1}
\label{pik1}
\end{equation}
The remaining contributions ${\bf \Pi}^D_1$ are of 
order $\nabla^2$ and will be neglected. The final expression
of the stress tensor in linear order of gradients is given by
collecting (\ref{pito}), (\ref{apito}) and (\ref{pik1})

\begin{eqnarray}
\frac{1}{D}{\rm tr}[{\bf \Pi}]&=&- \left[\gamma m n^2
\left[\frac{A_2}{2D}+\frac{(D+2)}{2D}B_2\right]
+\frac{k_BT}{\gamma D [A_0+B_0]}
\right]\nabla\!\cdot\!{\bf u} + nk_BT
\nonumber\\
{\overline {\bf \Pi}}^S&=&
-\left[
\gamma m n^2\left[\frac{A_2}{2}+B_2\right]
+\frac{k_BT}{\gamma[A_0+B_0]}\right]
\overline{\nabla {\bf u}}^S
\nonumber\\
{\bf \Pi}^A&=&-\gamma m n^2\frac{A_2}{2}
\left[\frac{1}{2}\nabla\times{\bf u}-{\bf \Omega}\right]
\label{stressfin}
\end{eqnarray}

Comparison of (\ref{phen}) and (\ref{stressfin}) allows to 
identify the viscosities as 

\begin{eqnarray}
\zeta&=&\left[
\gamma m n^2
\left[\frac{A_2}{2D}+\frac{(D+2)}{2D}B_2\right]
+\frac{k_BT}{\gamma D[A_0+B_0]}
\right]
\nonumber\\
\eta&=&\frac{1}{2}\left[
\gamma m n^2\left[\frac{A_2}{2}+B_2\right]
+\frac{k_BT}{\gamma[A_0+B_0]}\right]
\nonumber\\
\eta_R&=&\gamma m n^2\frac{A_2}{2}
\label{transport}
\end{eqnarray}
In order to compare this expressions with those obtained by Marsh {\em
et al.} in Ref. \cite{mar97}, we should note that for the original
DPD algorithm we have

\begin{eqnarray}
A(r)&=&0
\nonumber\\
B(r)&=&\omega(R)
\label{comparison}
\end{eqnarray}
Simple substitution of (\ref{comparison}) into (\ref{transport}) shows
that the transport coefficient (\ref{transport}) coincide with those
provided in Ref. \cite{mar97}.

\subsection{Transport equations}
Substitution of the stress tensor ${\bf \Pi}={\rm tr}[{\bf \Pi}]{\bf
1}/D +{\overline {\bf \Pi}}^S +{\bf \Pi}^A$ (\ref{phen}) into the momentum
balance equation (\ref{balance}) produces the Navier-Stokes equations
for a fluid with spin \cite{gro62} ($D=3$),

\begin{equation}
\rho\frac{d}{dt}{\bf u}=-\nabla p+\nabla\!\cdot\!(2\eta \nabla {\bf u}^S)
+\nabla(\zeta-2\eta/3)\nabla\!\cdot\!{\bf u}
+\nabla\times\left[2\eta_R({\bf \Omega}-\frac{1}{2}\nabla\times{\bf
u})\right]
\label{NSE}
\end{equation}
where we have used the substantial derivative $d/dt=\partial_t+{\bf
u}\!\cdot\!\nabla$.  The last term in (\ref{NSE}) is the gradient of the
antisymmetric part of the stress tensor and describes the effect of the
spin on the momentum transport.

On the other hand by neglecting the term ${\bf \Phi}$ in the angular
momentum balance equation (\ref{bj}) \cite{gro62}  we obtain
\begin{equation}
\partial_t {\bf J}=-\nabla\left[{\bf J}{\bf v}+{\bf r}\times {\bf \Pi}\right]
\label{bjbis}
\end{equation}
which in combination with (\ref{L}) produces the following
balance equation for the spin density

\begin{equation}
\partial_t {\bf S}=- \nabla [{\bf S}{\bf u}]-2{\bf \Pi}^A
\label{S}
\end{equation}
which, implies the following dynamic equation for the angular
velocity
\begin{equation}
nI\frac{d}{dt}{\bf \Omega}=-2{\bf \Pi}^A
\label{Ombis}
\end{equation}
Substitution of ${\bf \Pi}^A$ in 
(\ref{phen}) into this equation gives
\begin{equation}
\frac{d}{dt}{\bf \Omega}=
-\frac{4\eta_R}{n I}
\left[{\bf\Omega}-\frac{1}{2}\nabla\times{\bf u}\right]
=-\frac{1}{\tau}
\left[{\bf\Omega}-\frac{1}{2}\nabla\times{\bf u}\right]
\label{om}
\end{equation}
The final closed set of equations for the hydrodynamic fields is given
by
Eqns. (\ref{NSE}), (\ref{om}), together with the continuity equation
\begin{equation}
\frac{d}{dt}\rho = -\rho\nabla\!\cdot\!{\bf u}
\label{masscont}
\end{equation}
and the equation of state

\begin{equation}
p=k_BT n = \frac{k_BT}{m}\rho\equiv c^2 \rho
\label{presion}
\end{equation}
where we have introduced the speed of sound $c$.

Eqn. (\ref{om}) shows that the spin relaxes towards the vorticity with
a relaxation time scale given by $\tau=n I/4\eta_R$ \cite{gro62}.  In the
model of this paper, substitution of $\eta_R$ in (\ref{transport}) gives
the following time scale

\begin{equation}
\tau = \frac{I}{2\gamma nmA_2}
\label{tau}
\end{equation}
\subsection{Summary of kinetic theory}
In summary, it has been shown in this section that the macroscopic
behavior of the fluid particle model is hydrodynamical and the mass,
momentum and angular momentum transport equations have been derived
(Eqns. (\ref{masscont}), (\ref{NSE}), and (\ref{om})). In this doing,
explicit expressions for the transport coefficients in terms of the
original model parameters have been obtained (Eqns. (\ref{transport})
and (\ref{tau})). The equations here cited are the main results of the
kinetic theory of the fluid particle model.

\section{Resolution issues of the fluid particle model}

Within the picture of the Voronoi coarse-graining sketched in section
II, it is possible to consider different levels of coarse-graining in
which the number of atomic particles within a Voronoi cell is
different. We expect that, provided that the number of atomic
particles within the cell is large enough, the description of the
hydrodynamic behavior will be more and more accurate as the number of
Voronoi cells increases. In other words, we expect to reach a
``continuum limit'' as the number density of {\em fluid} particles
goes to infinity. The discussion resembles that of the resolution in
the numerical solution of partial differential equations. Actually,
the resemblance can be made more accurate by comparing the structure
of the equations of motion of the fluid particle model with those of
smoothed particle dynamics. Smoothed particle dynamics is a Lagrangian
discretization of the continuum equations of hydrodynamics that allows
to interpret the nodes of the grid in terms of ``smoothed
particles''. 

For the case in which there is no coupling between the Navier-Stokes
equation and the energy equation (the pressure does not depends on the
temperature, for example), Takeda {\em et al.}  \cite{tak94} propose a
discretization of the Navier-Stokes equations that produce equation of
motions for the smoothed particles that corresponds exactly in
structure with the postulated equations of motion of the fluid
particle model in this paper. The correspondence is

\begin{eqnarray}
V(r)&=&2\frac{p_0}{mn_0^2}W(r)
\nonumber\\
\gamma  A(r)&=&\frac{1}{mn_0^2}
\left[\eta W''(r)
+\left[2\eta+\left(\zeta+\frac{\eta}{3}\right)\right]
\frac{W'(r)}{r}\right]
\nonumber\\
\gamma  B(r)&=&\frac{1}{mn_0^2}
\left(\zeta+\frac{\eta}{3}\right)\left[ W''(r)-
\frac{W'(r)}{r}\right]
\label{corres}
\end{eqnarray}
where, $p_0,n_0$ are the equilibrium pressure and number density,
respectively, and $W(r)$ is the weight function used in the
discretization of the Navier-Stokes equation (the assumption that the
density of all particles is almost constant has been taken). 

In this respect, the fluid particle model postulated in this paper is
simply the smoothed particle dynamics with two additional bonus: 1)
thermal noise is introduced consistently (that is, the fluid particle
model can be interpreted as a Lagrangian discretization of the
non-linear fluctuating hydrodynamic equations), and 2) the angular
momentum is conserved exactly in the fluid particle model, in contrast
with the smoothed particle dynamics model. The first bonus allows to
apply smoothed particle dynamics to microhydrodynamic problems as
those appearing in complex fluids where Brownian fluctuations are due
to the fluctuating hydrodynamic environment. It can be also useful in
studying the effect of thermal fluctuations near convective
instabilities and, in general, in the study of non-equilibrium thermal
fluctuations in hydrodynamic systems. The actual relevance of the
second bonus will be discussed later.

The comparison of SPD with the fluid particle model points out to an
inconsistency that appears when using some particular selections for
the weight function like the Lucy weight function \cite{kum95} or a
Gaussian weight function \cite{tak94}. In these cases, it is easily
seen that the function $A(r)$ can become negative for certain values
of $r$. This is unacceptable in view of Eqn. (\ref{AB}). From a
physical point of view this means that if two particles are at a
distance such that $A(r)$ is negative, then the viscous forces will
try to {\em increase} its relative velocities!

In the derivation of the SPD model \cite{mon92},\cite{tak94} it
becomes apparent that the weight function $W(r)$ must be normalized to
unity, in order to have correct discrete (Monte-Carlo) approximations
for integrals.  If $W(r)$ is normalized to unity, then one expects
that by increasing the number density of smoothed particles one is
increasing the numerical resolution of the simulation. The
normalization implies that as the range of the weight function
decreases with higher resolution, its height increases and, in the
limit of infinite resolution it becomes a Dirac delta function
$W(r)\rightarrow \delta(r)$. Because the weight function is steeper
when the resolution is higher, the time step used in the SPD model has
to be reduced as the resolution increases. This is also encountered in
any finite difference algorithm for solution of partial differential
equations in order to maintain stability.

Let us investigate the effect of the resolution effects on the macroscopic
parameters defining the fluid on hydrodynamic scales and which have
been computed by means of the kinetic theory in the previous section.
The parameters that characterize the evolution of the velocity field
are, as can be appreciated from (\ref{NSE}), the speed of sound
defined in (\ref{presion}) and the {\em kinematic} viscosities 
defined by $\nu=\eta/\rho$, $\nu_b=\zeta/\rho$, and $\nu_r=\eta_R/\rho$.
From (\ref{transport}) they have the form

\begin{eqnarray}
\nu_b =&=&\left[
\gamma  n
\left[\frac{\overline{A}_2}{2D}+\frac{(D+2)}{2D}\overline{B}_2\right]
+c^2\frac{1}{\gamma D n[\overline{A}_0+\overline{B}_0]}
\right]
\nonumber\\
\nu&=&\frac{1}{2}\left[
\gamma n\left[\frac{\overline{A}_2}{2}+\overline{B}_2\right]
+c^2\frac{1}{\gamma n[\overline{A}_0+\overline{B}_0]}\right]
\nonumber\\
\nu_R&=&\gamma  n\frac{\overline{A}_2}{2}
\label{kintrans}
\end{eqnarray}
We are assuming, for the sake of the argument, that $n=n_0$, that
is, the density field is constant. The conclusions, however, are
valid in the compressible case also.

Let us focus first on the dimensionless functions $A(r),B(r)$ that
determine the range of the dissipative and random forces. We expect
that the clusters interact only with their neighbors, which are a
typical distance $\lambda$ apart. Therefore, these functions will be
of the form

\begin{eqnarray}
A(r)&=&a(r/\lambda)
\nonumber\\
B(r)&=&b(r/\lambda)
\end{eqnarray}
where $a,b$ are functions that do not depend explicitly on
$\lambda$. This ensures that as the resolution is increased, the range
of the force decreases, and this has the computationally appealing
feature that the interaction between fluid particles remains always
local. By using these scaling function and after a change to the
dimensionless variable ${\bf x}\equiv {\bf r}/\lambda$, we have

\begin{eqnarray}
\overline{A}_0&=&\frac{a_0}{n_0}
\nonumber\\
\overline{B}_0&=&\frac{b_0}{n_0}
\nonumber\\
\overline{A}_2&=&\frac{a_2}{n_0}\lambda^2
\nonumber\\
\overline{B}_2&=&\frac{b_2}{n_0}\lambda^2
\label{extract}
\end{eqnarray}
where the dimensionless coefficients are given by
\begin{eqnarray}
a_0&=&\int a(x)d^D{\bf x}
\nonumber\\
b_0&=&\int b(x)d^D{\bf x}
\nonumber\\
a_2&=&\int x^2 a(x)d^D{\bf x}
\nonumber\\
b_2&=&\int x^2 b(x)d^D{\bf x}
\label{norm2}
\end{eqnarray}
and do not depend on the resolution. By using (\ref{extract}) into
(\ref{kintrans}) we obtain

\begin{eqnarray}
\nu_b&=&\left[
\gamma \lambda^2\left[\frac{a_2}{2D}+\frac{(D+2)}{2D}b_2\right]
+c^2\frac{1}{D \gamma(a_0+b_0)}
\right]
\nonumber\\
\nu&=&\frac{1}{2}\left[
\gamma \lambda^2\left[\frac{a_2}{2}+b_2\right]
+c^2\frac{1}{\gamma(a_0+b_0)}\right]
\nonumber\\
\nu_R&=&\gamma  \frac{\lambda^2}{2}a_2
\label{kintrans2}
\end{eqnarray}
We observe that all the dependence on the resolution ($\lambda$ or
$n_0$) has been made explicit. In the limit of high resolution
($\lambda\rightarrow0$ or $n_0\rightarrow\infty$) the only
contribution to the bulk $\nu_b$ an shear $\nu$ viscosities comes from
the kinetic contribution that depends linearly on the
temperature. This means that at zero temperature the system would not
display any viscosity in the limit of high resolution. We find this
behavior undesirable and we are lead to the conclusion that the
friction coefficient $\gamma$ must depend on $\lambda$. In particular,
if we define $\tilde{\gamma}\equiv \gamma \lambda^2$ (which has
dimensions of a kinematic viscosity) and assume that $\tilde{\gamma}$
remains constant as the resolution varies, we will have

\begin{eqnarray}
\nu_b&=&\left[
\tilde{\gamma}\left[\frac{a_2}{2D}+\frac{(D+2)}{2D}b_2\right]
+c^2\frac{\lambda^2}{D\tilde{ \gamma}}
\right]
\nonumber\\
\nu&=&\frac{1}{2}\left[
\tilde{\gamma}\left[\frac{a_2}{2}+b_2\right]
+c^2\frac{\lambda^2}{\tilde{\gamma}}\right]
\nonumber\\
\nu_R&=&  \frac{\tilde{\gamma}}{2}a_2
\label{kintrans3}
\end{eqnarray}
where the normalization $a_0=b_0=1$ has been used as in the original
DPD algorithm \cite{hoo92}. The relaxation time (\ref{tau}) will take the form

\begin{equation}
\tau=\frac{I}{2m}\frac{\lambda^2}{\tilde{\gamma} a_2}
\label{tau3}
\end{equation}
In this way, in the limit of high resolution ($\lambda\rightarrow 0$)
the viscosities are given essentially by $\tilde{\gamma}$ and the
relaxation time goes to zero (note that the moment of inertia $I$ must
be of the form $\propto m\times \lambda^2$ so $\tau$ must tend to zero
very fast). In the high resolution limit the spin becomes equal to the
vorticity in a short time scale. The spin becomes a slaved variable
and can be dropped from the description. Note also that in these
situation the last term in the Navier-Stokes equation with spin
(\ref{NSE}) vanishes and one recovers the actual Navier-Stokes
equation. This explains why, in SPD, the violation of angular momentum
does not poses a serious problem {\em for sufficiently high
resolutions} \cite{tak94}. If low resolutions are to be used in
problems where the correct transfer of angular momentum is relevant
(like in rotational diffusion of concentrated colloidal suspensions,
for example), then the use of spin might suppose a real advantage.

We have arrived at the conclusion that in order have a well-defined
continuum limit the friction coefficient $\gamma$ must increase as the
resolution increases. This can be understood physically in the
following way. The number of particles in between two reference fluid
particles at a given distance of each other increases as the
resolution increases. If we require that the viscous interaction
between these two reference particles must remain the same as the
resolution increases, the mediating particles must interact stronger
in order to transmit the same response between the two reference
particles. From a mathematical point of view, the $\lambda^2$ factor
can be interpreted as the ``lattice spacing'' that is lacking in the
original equations and that would be present in a numerical
discretization of a {\em second} order derivative term. Preliminary
simulation results for the DPD model ($A(r)=0$) with energy conservation
\cite{esp-ener97} shows that the correct continuum limit is obtained 
when the model parameter equivalent to $\gamma$ increases with
$\lambda^2$ \cite{rip97}.

We would like to comment finally on an apparent inconsistency between
SPD and the fluid particle model which is summarized as follows: if
one discretizes the hydrodynamic equations on a set of points and then
constructs the kinetic theory of these points, one would expect that
the computed transport coefficients would coincide with the input
transport coefficients of the hydrodynamic equations.  If one naively
uses the results (\ref{corres}) in the calculation of the transport
coefficients in (\ref{transport}), one arrives at an inconsistent
result. The viscosities computed through the kinetic theory
(\ref{transport}) do not coincide with the input values. This could be
traced back to the fact that the kinetic theory for the fluid particle
model has been developed in the limit where no conservative forces are
present, whereas the pressure term in the hydrodynamic equations (even
for an ideal gas!) produces a conservative term given by the first
equation in (\ref{corres}) in SPD. The kinetic theory with
conservative forces is a bit more involved but the modifications can
be summarized simply. The molecular chaos assumption (\ref{chaos}) now
involves the pair distribution function (which in the absence of
conservative forces is equal to 1). This means that the parameters
$\overline{A}_2,\overline{B}_2$ appearing in the transport
coefficients will be modified by the presence of the pair distribution
function within the integral defining these parameters. Also a new
contribution to the transport coefficients arises due to the
conservative forces. It is an open question whether these modified
transport coefficients due to conservative forces do coincide with the
input transport coefficients.  The opposite case could also be
possible simply due to the fact that the discretization procedure in
SPD may induce ``artificial viscosities'' in the language of numerical
resolution of the hydrodynamic partial differential equations.

The fluid particle model is a consistent model by itself, without
having to resort to the smoothed particle model for its
validity. Actually, the fluid particle model, together with the
kinetic theory developed in this paper has its advantages with respect
to SPD: precise predictions can be made from the initial model
parameters about the transport properties of the fluid. In this way,
to obtain a prescribed fluid of known transport properties, one simply
adjust the model parameters according to the formulae of kinetic
theory (slight errors stemming from the failure of the molecular chaos
assumption might play a minor role \cite{mar97}). In SPD, on the
contrary, the only way to specify the fluid is through the input
transport coefficients in the original hydrodynamic equations. The
discretization procedure then produces a ``fluid'' whose transport
properties do not in general correspond with those of the fluid
intended to be modeled, and there is no systematic control on the
appearance of artificial viscosities.

It is apparent that this whole discussion can be applied to the DPD
model, which is a particular case of the fluid particle model, and for
which a kinetic theory has been formulated previously in
Ref. \cite{mar97}. One of the main motivations for introducing shear
forces between dissipative particles into the original algorithm of
DPD was the identification of the following elementary motion between
dissipative particles that produces no force in that algorithm. Let us
focus on two neigbouring dissipative particles, the first one at rest
at the origin and the second one orbiting in a circumference around
the first one. This relative motion produces no force in DPD because
the relative approaching velocity is exactly zero. Nevertheless, on
simple physical grounds one expect that the motion of the second
particle must drag in some way the first particle. This is taken into
account through the shear forces in the fluid particle model presented
in this paper.  We note, however that this relative motion might
produce a drag even in the original DPD algorithm {\em if many DPD
particles are involved simultaneously}. The same is true for a purely
conservative molecular dynamics simulation. The point is, of course,
that the effect is already captured with a much smaller number of
particles in the fluid particle model.

\section*{Acknowledgments}
I am very much indebted to M. Serrano, M. Ripoll, M.A. Rubio, and I.
Z\'{u}\~niga for their comments and suggestions during the elaboration
of this work. I express here my gratitude to M. Ernst and Wm. G.
Hoover for the illuminating correspondence and to C. Marsh and G.
Backx for making available their work previous to publication. This
work has been partially supported by a DGICYT Project No PB94-0382 and
by E.C. Contract ERB-CHRXCT-940546.


\section*{Appendix}
The derivation of the FPE is best achieved by considering the
differential of an arbitrary function $f$ to second order \cite{gar83}

\begin{eqnarray}
d f &=& \sum_i d{\bf r}_i\frac{\partial f}{\partial {\bf r}_i}
+d{\bf v}_i\frac{\partial f}{\partial {\bf v}_i}
+d{\bf \omega}_i\frac{\partial f}{\partial {\bf \omega}_i}
\nonumber\\
&+&
\frac{1}{2}\sum_{ij}
d{\bf v}_id{\bf v}_j\
\frac{\partial^2 f}{\partial {\bf v}_i\partial {\bf v}_j}
+
d{\bf v}_id{\bf \omega}_j 
\frac{\partial^2 f}{\partial {\bf v}_i\partial {\bf \omega}_j}
+
d{\bf \omega}_id{\bf v}_j
\frac{\partial^2 f}{\partial {\bf \omega}_i\partial {\bf v}_j}
+
d{\bf \omega}_id{\bf \omega}_j
\frac{\partial^2 f}{\partial {\bf \omega}_i\partial {\bf \omega}_j}
\label{dif}
\end{eqnarray}
One then substitutes the SDE's (\ref{sdea}) and uses the Ito
stochastic rules (\ref{ran3}) keeping terms up to order $dt$ (the
cross terms involving positions have been neglected in (\ref{dif}) on
account of the fact that $d{\bf r}$ is already of order $dt$). Then
after averaging with respect to the distribution function
$\rho(r,v,\omega;t)$, one performs a partial integration and uses the
fact that $f$ is arbitrary, to obtain the Fokker-Planck equation in
the form

\begin{equation}
\partial_t \rho(r,v,\omega;t)=\left[L^C+L^T+L^R\right]\rho(r,v,\omega;t)
\label{fp1app}
\end{equation}
where we have defined the operators
\begin{eqnarray}
L^C
&\equiv&-\left[\sum_i{\bf v}_i\frac{\partial }{\partial {\bf r}_i}
+\sum_{i,j\neq i}\frac{1}{m}{\bf F}^C_{ij}
\frac{\partial}{\partial{\bf v}_i}\right]
\nonumber \\
L^T&\equiv&
\sum_{i,j\neq i}\frac{\partial}{\partial {\bf v}_i}
\left[
-\frac{1}{m}\left({\bf F}^T_{ij}+{\bf F}^R_{ij}\right)
+\frac{1}{2}\frac{\partial}{\partial {\bf v}_j}
\sum_{i'j'}\frac{1}{dt}d{\tilde{\bf v}}_{ii'}d{\tilde{\bf v}}_{jj'}\right.
\nonumber \\
&-&\left.\frac{1}{2}\frac{\partial}{\partial {\bf \omega}_j}\frac{m}{I}
\sum_{i'j'}\frac{1}{dt}d{\tilde{\bf v}}_{ii'}
\left(\frac{{\bf r}_{jj'}}{2}\times d{\tilde{\bf v}}_{jj'}\right)
\right]
\nonumber \\
L^R&\equiv&\frac{m}{I}
\sum_{i,j\neq i}\frac{\partial}{\partial {\bf \omega}_i}
\left[\frac{1}{m}
\frac{{\bf r}_{ij}}{2}\times\left({\bf F}^T_{ij}+{\bf F}^R_{ij}\right)
-\frac{1}{2}\frac{\partial}{\partial {\bf v}_j}
\sum_{i'j'}\frac{1}{dt}
\left(\frac{{\bf r}_{ii'}}{2}
\times d{\tilde{\bf v}}_{ii'}\right)d{\tilde{\bf v}}_{jj'}\right.
\nonumber \\
&+&\left.\frac{1}{2}\frac{\partial}{\partial {\bf \omega}_j}
\frac{m}{I}
\sum_{i'j'}\frac{1}{dt}
\left(\frac{{\bf r}_{ii'}}{2}\times d{\tilde{\bf v}}_{ii'}\right)
\left(\frac{{\bf r}_{jj'}}{2}\times d{\tilde{\bf v}}_{jj'}\right)
\right]
\label{fp2}
\end{eqnarray}
The operator $L^C$ is the usual Liouville operator of a Hamiltonian
system interacting with conservative forces ${\bf F}^C$. 
We need to arrange a bit the operators $L^T$ and $L^R$ by using the
Ito rules (\ref{ran4})

\begin{eqnarray}
\frac{1}{dt}d{\tilde{\bf v}}^\mu_{ii'}d{\tilde{\bf v}}^\nu_{jj'}
&=&
\sigma^2\left[\frac{1}{2}\left[{\tilde A}(r_{ii'}){\tilde A}(r_{jj'})
+
{\tilde C}(r_{ii'}){\tilde C}(r_{jj'})\right]
\delta^{\mu\nu}{\bf e}_{ii'}\!\cdot\!{\bf e}_{jj'}
\right.
\nonumber\\
&+&
\frac{1}{2}
\left[{\tilde A}(r_{ii'}){\tilde A}(r_{jj'})
-{\tilde C}(r_{ii'}){\tilde C}(r_{jj'})\right]
{\bf e}^\nu_{ii'}{\bf e}^\mu_{jj'}
\nonumber\\
&+&
\left.\frac{1}{D}\left[
{\tilde B}(r_{ii'}){\tilde B}(r_{jj'})-{\tilde A}(r_{ii'}){\tilde A}(r_{jj'})\right]
{\bf e}^\mu_{ii'}{\bf e}^\nu_{jj'}\right]
\nonumber\\
&\times&
\left[\delta_{ij}\delta_{i'j'}+\delta_{ij'}\delta_{ji'}\right]
\nonumber\\
&\equiv&\sigma^2
{\bf T}^{\mu\nu}_{ii'jj'}
\left[\delta_{ij}\delta_{i'j'}+\delta_{ij'}\delta_{ji'}\right]
\label{dvdv}
\end{eqnarray}
The second order tensor ${\bf T}^{\mu\nu}_{ii'jj'}$ satisfies
\begin{equation}
  {\bf T}^{\mu\nu}_{ijij}
= {\bf T}^{\nu\mu}_{ijij}
=-{\bf T}^{\mu\nu}_{ijji}
\label{ts}
\end{equation}
If we define
\begin{equation}
{\bf T}_{ij}\equiv{\bf T}_{ijij}=
\frac{1}{2}\left[{\tilde A}^2(r_{ij})
+{\tilde C}^2(r_{ij})\right]{\bf 1}
+\left[\left(\frac{1}{2}-\frac{1}{D}\right)
{\tilde A}^2(r_{ij})+\frac{1}{D}{\tilde B}^2(r_{ij})
-\frac{1}{2}{\tilde C}^2(r_{ij})\right]
{\bf e}_{ij}{\bf e}_{ij}\label{tij}
\end{equation}
then the following identities are obtained

\begin{eqnarray}
\frac{1}{2}\sum_{ij}
\frac{\partial}{\partial{\bf v}_i}\frac{\partial}{\partial{\bf v}_j}
\sum_{i'j'}\frac{1}{dt}d{\tilde{\bf v}}_{ii'}d{\tilde{\bf v}}_{jj'}
&=&
\frac{\sigma^2}{2}\sum_{ij}
\frac{\partial}{\partial{\bf v}_i}\!\cdot\!{\bf T}_{ij}\!\cdot\!
\left[\frac{\partial}{\partial{\bf v}_i}-\frac{\partial}{\partial{\bf v}_j}\right]
\nonumber \\
-\frac{1}{2}\sum_{ij}
\frac{\partial}{\partial{\bf v}_i}\frac{\partial}{\partial{\bf \omega}_j}
\sum_{i'j'}\frac{1}{dt}d{\tilde{\bf v}}_{ii'}
\left(\frac{{\bf r}_{jj'}}{2}\times d{\tilde{\bf v}}_{jj'} \right)
&=&
\frac{\sigma^2}{2} \sum_{ij}
\frac{\partial}{\partial{\bf v}_i} 
\!\cdot\!{\bf T}_{ij}\!\cdot\!\left(\frac{{\bf r}_{ij}}{2}\times
\left[\frac{\partial}{\partial{\bf \omega}_i}
+\frac{\partial}{\partial{\bf \omega}_j}\right]\right)
\nonumber \\
-\frac{1}{2}\sum_{ij}
\frac{\partial}{\partial{\bf \omega}_i}\frac{\partial}{\partial{\bf v}_j}
\sum_{i'j'}\frac{1}{dt}\left(\frac{{\bf r}_{ii'}}{2}
\times d{\tilde{\bf v}}_{ii'}\right)
d{\tilde{\bf v}}_{jj'}
&=&
-\frac{\sigma^2}{2} \sum_{ij}
\frac{\partial}{\partial{\bf \omega}_i}\!\cdot\!
\left(\frac{{\bf r}_{ij}}{2}\times
 {\bf T}_{ij}\!\cdot\!
\left[\frac{\partial}{\partial{\bf v}_i}
-\frac{\partial}{\partial{\bf v}_j}\right]\right)
\nonumber \\
\frac{1}{2}\sum_{ij}
\frac{\partial}{\partial{\bf \omega}_i}\frac{\partial}{\partial{\bf \omega}_j}
\sum_{i'j'}\frac{1}{dt}\left(\frac{{\bf r}_{ii'}}{2} 
\times d{\tilde{\bf v}}_{ii'}\right) 
\left(\frac{{\bf r}_{jj'}}{2}\times d{\tilde{\bf v}}_{jj'}\right)
&=&
-\frac{\sigma^2}{2}\sum_{ij}
\frac{\partial}{\partial{\bf \omega}_i}\!\cdot\!
\left(\frac{{\bf r}_{ij}}{2}\times
{\bf T}_{ij}\!\cdot\!\left(\frac{{\bf r}_{ij}}{2}\times 
\left[\frac{\partial}{\partial{\bf \omega}_i}
+\frac{\partial}{\partial{\bf \omega}_j}\right]\right)\right)
\label{terms}
\end{eqnarray}
By using these results into (\ref{fp2}) the operators take the
following compact form

\begin{eqnarray}
L^T&=&
\sum_{i,j\neq i}\frac{\partial}{\partial {\bf v}_i}
\!\cdot\!\left[{\bf L}^T_{ij}+{\bf L}^R_{ij}\right]
\nonumber \\
L^R&=&-\frac{m}{I}
\sum_{i,j\neq i}\frac{\partial}{\partial {\bf \omega}_i}
\!\cdot\!\left(\frac{{\bf r}_{ij}}{2}\times
\left[{\bf L}^T_{ij}+{\bf L}^R_{ij}\right]\right)
\label{fp3app}
\end{eqnarray}
where we have introduced the vector operators
\begin{eqnarray}
{\bf L}^T_{ij}
&\equiv&-\frac{1}{m}{\bf F}^T_{ij}
+\frac{\sigma^2}{2}{\bf T}_{ij}\!\cdot\!
\left[\frac{\partial}{\partial{\bf v}_i}-\frac{\partial}{\partial{\bf v}_j}\right]
\nonumber\\
{\bf L}^R_{ij}&\equiv&
-\frac{1}{m}{\bf F}^R_{ij}
+\frac{m}{I}\frac{\sigma^2}{2}{\bf T}_{ij}
\!\cdot\!
\left( \frac{{\bf r}_{ij}}{2}\times \left[\frac{\partial}{\partial{\bf \omega}_i}
+\frac{\partial}{\partial{\bf \omega}_j}\right]\right)
\label{vecoperapp}
\end{eqnarray}

\end{document}